\documentclass[12pt,final]{iopart}
\usepackage{color,cite}
\usepackage{iopams,graphicx}
\usepackage{braket}
\usepackage{dsfont}

\newcommand{\llangle}{\left \langle}
\newcommand{\rrangle}{\right \rangle}

\newenvironment{remark}[1][Remark:]{\begin{trivlist}
\item[\hskip \labelsep {\bfseries #1}]}{\end{trivlist}}

\begin{document}

\title[]{Correlations of the density and of the current in non-equilibrium diffusive systems}
\author{Tridib Sadhu$^{1}$ and Bernard Derrida$^{2}$}
\vspace{1pc}
\address{$^1$ Philippe Meyer Institute for Theoretical Physics, Physics Department, \\Ecole Normale Sup\'{e}rieure, 24 rue Lhomond, 75231 Paris Cedex 05 - France.}
\ead{tridib.sadhu@lpt.ens.fr}
\vspace{1pc}
\address{$^2$ Coll\`{e}ge de France, 11 place Marcelin Berthelot, 75231 Paris Cedex 05 - France.}
\ead{derrida@lps.ens.fr}
\begin{abstract}
We use fluctuating hydrodynamics to analyze the dynamical properties in the non-equilibrium steady state of a diffusive system coupled with reservoirs. We derive the two-time correlations of the density and of the current in the hydrodynamic limit in terms of the diffusivity and the mobility. Within this hydrodynamic framework we discuss a generalization of the fluctuation dissipation relation in a non-equilibrium steady state where the response function is expressed in terms of the two-time correlations. We compare our results to an exact solution of the symmetric exclusion process. This exact solution also allows one to directly verify the fluctuating hydrodynamics equation.
\end{abstract}

\date{\today}
\vspace{2pc}

\submitto{J Stat Mech}
\maketitle

\section{Introduction}
Understanding the steady state properties of non-equilibrium systems is an important topic in modern statistical physics.
For interacting particle systems a general approach is based on a fluctuating hydrodynamics description \cite{Spohn1991,KIPNISLANDIM}. Such a description is the starting point of a recently developed macroscopic fluctuation theory \cite{Bertini2014,Jona-Lasinio2010,Bertini2009,Derrida2007,Bertini2001} which has proved useful to calculate large deviation functions of diffusive systems. While a large amount of results have been obtained based on the fluctuating hydrodynamics approach, they are mostly about one-time statistics \cite{Spohn1983,Spohn1991,KIPNISLANDIM,GARRIDO,Gerschenfeld2009,Krapivsky2012,Tailleur2007,Tailleur2008,Eyink1990,
Lecomte2010,DeMasi1982,Akkermans2013,Hurtado2011,Bodineau2010,Bernardin2008,Bunin2012}. In this work, we use this framework to see how some other dynamical properties in a non-equilibrium steady state, namely multi-time correlations, spectral distribution, and linear-response can be calculated.

We consider the non-equilibrium steady state of a one-dimensional system coupled with reservoirs at two ends. It may be a system of interacting particles coupled with reservoirs at different chemical potentials, or a thermal conducting rod coupled with heat baths at different temperatures. One uses fluctuating hydrodynamics to describe such a system in the large system size $L$ limit; in systems where diffusion is the transport mechanism, one defines hydrodynamics coordinates $x=\frac{X}{L}$ and $\tau=\frac{t}{L^2}$ where $X$ and $t$ are position and time, respectively. The time evolution of the system is described in terms of a hydrodynamic density $\rho(x,\tau)$ and a hydrodynamic current $J(x,\tau)$ which satisfy \cite{Spohn1991,KIPNISLANDIM}
\begin{equation}
\fl \qquad \partial_{\tau}\rho(x,\tau)=-\partial_x J(x,\tau)\quad \textrm{with}\quad J(x,\tau)+D(\rho(x,\tau))\partial_x\rho(x,\tau)=\eta(x,\tau) 
\label{eq:fhd 0}
\end{equation}
where $D(\rho)$ is the diffusivity. What \eref{eq:fhd 0} tells us is that on average the current follows Fick's law ($J=-D(\rho)\partial_x\rho $) and the fluctuations are modeled by a Gaussian random noise  $\eta(x,\tau)$ with a covariance
\begin{equation}
\llangle \eta(x,\tau)\eta(y,\tau^{\prime})\rrangle=\frac{1}{L}\sigma(\rho(x,\tau))\,\delta(x-y)\,\delta(\tau-\tau^{\prime})
\label{eq:etaeta cov}
\end{equation}
where $\sigma(\rho)$ is the mobility.

In this paper, we use \eref{eq:fhd 0}-\eref{eq:etaeta cov} to derive the two-time correlations $\llangle \rho(x,\tau)\rho(y,\tau^{\prime}) \rrangle_c$ and $\llangle J(x,\tau)J(y,\tau^{\prime})\rrangle_c$, and to obtain the correlations of the integrated current $q(x,\tau)$ which is defined as the time integral of $J(x,\tau)$. We will show in particular that
\begin{equation}
\fl \llangle q(x,\tau)q(x,\tau^{\prime}) \rrangle_c\simeq \cases{\frac{\min(\tau,\tau^{\prime})}{L}\int_{0}^{1}dz\sigma(\overline{\rho}(z)) & for $\tau,\>\tau^{\prime}\gg 1$ \cr\frac{1}{L}\left[ \sqrt{\tau} +\sqrt{\tau^{\prime}} -\sqrt{\vert\tau-\tau^{\prime}\vert} \right]\frac{\sigma(\overline{\rho}(x))}{2\sqrt{\pi\,D(\overline{\rho}(x))}} & for  $\tau,\>\tau^{\prime}\ll 1$}
\end{equation}
where $\overline{\rho}(x)$ is the average density in the steady state.
At large times the asymptotic behavior corresponds to a standard Brownian motion with known \cite{Bodineau2004} variance whereas at short time the covariance is that of a fractional Brownian motion \cite{Mandelbrot1968}. One can notice that the short time behavior also gives the correlation of the integrated current of an infinite system at equilibrium. (One may infer the time dependence from the correlation of the height fluctuations in the Edwards-Wilkinson interface growth \cite{KrugKallabis1997}.)

Our approach is valid for a general diffusive system in a non-equilibrium steady state for which the microscopic details are embedded in the two transport coefficients $D(\rho)$ and $\sigma(\rho)$ which appear in (\ref{eq:fhd 0},\ref{eq:etaeta cov}). We will compare our results with an exact solution of the symmetric exclusion process which corresponds to $D(\rho)=1$ and $\sigma(\rho)=2\rho(1-\rho)$ \cite{Bodineau2004,Derrida2007}. This is a system of diffusing particles on a lattice with simple exclusion interactions among them. Starting with microscopic dynamics we will derive an explicit formula of the two-time correlations $\llangle \tau_i(t)\tau_j(t^{\prime})\rrangle_c$ and $\llangle \mathcal{J}_i(t)\mathcal{J}_j(t^{\prime})\rrangle_c$ for the occupation number $\tau_i(t)=0,~ 1$ of the site $i$ and for the particle current $\mathcal{J}_i(t)$ across a bond $(i,i+1)$. We show that, in the hydrodynamic limit, the formulas lead to their corresponding results for $\llangle \rho(x,\tau)\rho(y,\tau^{\prime})\rrangle_c$  and $\llangle J(x,\tau)J(y,\tau^{\prime})\rrangle_c $.
These exact calculations allow us to directly verify the fluctuating hydrodynamics equation \eref{eq:fhd 0}. If we define $Y_i(t)=\mathcal{J}_i(t)-\left[\tau_i(t)-\tau_{i+1}(t)\right]$, \textit{i.e.}, $Y_i(t)$ is the difference between the actual current and the ``microscopic" Fick's, we shall show that $\llangle Y_i(t)\rrangle=0$ and the covariance 
\begin{equation}
\llangle Y_i(t)Y_j(s)\rrangle=\delta(t-s)\delta_{i,j}\left\{2\llangle \tau_i(t)\rrangle\left[1-\llangle\tau_{i}(t)\rrangle \right]+\Or\left(\frac{1}{L}\right)\right\}
\label{eq:YY}
\end{equation}
which is the microscopic version of \eref{eq:etaeta cov} with the mobility $\sigma(\rho)$ for the symmetric exclusion process.

In the last part of this paper, we use the fluctuating hydrodynamics to see how the linear response is modified in a non-equilibrium steady state. In equilibrium, the fluctuation dissipation theorem gives a simple relation between the response function to a small perturbation and the correlation functions in the unperturbed system \cite{Kubo1966,Ruelle2009}. Many recent works \cite{Seifert2010,Baiesi2009,Baiesi20092,Maes2010,Maes20102,Diezemann2005,Prost2009,Chetrite2008,Chatelain2003,Corberi2007
,Cugliandolo1994,HANGGI1982,Agarwal1972} have discussed extensions of this relation to non-equilibrium systems.

Here we consider a perturbation produced by a small external field $h(x,\tau)$ in the hydrodynamic current so that the formula  in \eref{eq:fhd 0} giving $J(x,\tau)$ is replaced by 
\begin{equation}
J_h(x,\tau)+D(\rho(x,\tau))\partial_x\rho(x,\tau)=\eta(x,\tau)+\sigma(\rho(x,\tau))h(x,\tau).
\label{eq:5}
\end{equation}
The change in the average density profile
\begin{equation}
\llangle \Delta \rho(x,\tau)\rrangle=\int_{-\infty}^{\tau}d\tau^{\prime}\int dy \,h(y,\tau^{\prime})R(x,\tau; y,\tau^{\prime})
\label{eq:R rho}
\end{equation}
where $R$ is the associated response function.
We will show that in the steady state
\begin{eqnarray}
\fl \qquad \frac{1}{L}R(x,\tau;y, \tau^{\prime})\simeq \frac{d}{d\tau^{\prime}}\llangle \rho(x,\tau)q(y,\tau^{\prime})\rrangle_c-\frac{d}{d\tau}\llangle q(y,\tau) \rho(x,\tau^{\prime})\rrangle_c\cr
 \qquad  +\frac{d}{dy}\left[D(\overline{\rho}(y))\llangle \rho(x,\tau)\rho(y,\tau^{\prime})\rrangle_c\right]- \frac{d}{dy}\left[D(\overline{\rho}(y)) \llangle\rho(y,\tau)\rho(x,\tau^{\prime})\rrangle_c\right]
\label{eq:Rq 4 0}
\end{eqnarray}
(the $\simeq$ sign means that higher order terms in $\frac{1}{L}$ are ignored.)
The equilibrium fluctuation dissipation relation is merely a special case of the formula \eref{eq:Rq 4 0} where the last two terms cancel each other and the first two terms are equal due to time reversal symmetry. Formula \eref{eq:Rq 4 0} can be seen as an explicit example of a general result in \cite{Baiesi2009,Baiesi20092}. 

This paper is organized as follows. In \sref{sec:fhd}, we discuss the fluctuating hydrodynamics and use it to calculate two-time correlations. An exact derivation of the correlations in the symmetric exclusion process is presented in \sref{sec:sep}. In \sref{sec:linear response}, we discuss the linear response relation.

\section{Fluctuating hydrodynamics \label{sec:fhd}}
It is straightforward to derive two-time correlations of the density, and of the current in terms of a Green's function within the framework of fluctuating hydrodynamics. One starts with the fluctuating hydrodynamics equation \eref{eq:fhd 0} where boundary condition comes from the density of the reservoirs leading to
\begin{equation}
\rho(0,\tau)=\rho_a \qquad  \textrm{and}\qquad  \rho(1,\tau)=\rho_b \qquad \textrm{at all times $\tau$.}
\label{eq:rho boundary}
\end{equation}
In the steady state, the average profile $\overline{\rho}(x)$ is obtained as the solution of
\begin{equation}
\partial_x\left[D(\overline{\rho}(x))\partial_x\overline{\rho}(x)\right]=0 \qquad \textrm{with \qquad $\overline{\rho}(0)=\rho_a$ \quad and \quad $\overline{\rho}(1)=\rho_b$ }
\label{eq:av profile}
\end{equation}

The density fluctuates around $\overline{\rho}(x)$ and we denote such a fluctuation as $r(x,\tau)=\rho(x,\tau)-\overline{\rho}(x)$.
To calculate the two-time correlation, it is sufficient to consider small fluctuations. Using \eref{eq:fhd 0} and keeping up to linear order in $r(x,\tau)$ one obtains
\begin{equation*}
\partial_{\tau} r(x,\tau)-\partial^2_{x} \left[D(\overline{\rho}(x))r(x,\tau)\right]=-\partial_x\eta(x,\tau)
\label{eq:r equation}
\end{equation*}
with the boundary conditions $r(0,\tau)=0$ and $r(1,\tau)=0$ which come from \eref{eq:rho boundary}.

The solution can be written after an integration by parts.
\begin{equation}
r(x,\tau)=\int_{-\infty}^{\tau}d\tau^{\prime}\int_0^1dz\,\eta(z,\tau^{\prime})\,\partial_{z}G(x, z,\tau-\tau^{\prime})
\label{eq:solution for r}
\end{equation}
where the Green's function is defined as the solution of
\begin{equation}
\partial_{\tau} G(x,y,\tau)-\partial^2_{x} \left[D(\overline{\rho}(x))G(x,y,\tau)\right]=0
\label{eq:G}
\end{equation}
with $G(x,y,0)=\delta(x-y)$ and a boundary condition that the Green's function vanishes at the spatial boundaries of the system:
\begin{eqnarray*}
G(0,y,\tau)=0; \qquad G(1,y,\tau)=0 \qquad \textrm{for all $y$} \cr
G(x,0,\tau)=0; \qquad G(x,1,\tau)=0 \qquad \textrm{for all $x$}
\end{eqnarray*}
A useful identity for the Green's function is
\begin{equation}
G(x,y,\tau+s)=\int dz G(x,z,\tau)G(z,y,s)
\label{eq:id G one}
\end{equation}
which allows one to verify that the Green's function satisfies also
\begin{equation}
\partial_{\tau} G(x,y,\tau)-D(\overline{\rho}(y))\partial^2_{y} G(x,y,\tau)=0
\label{eq:G2}
\end{equation}
which will be used later in the calculation.

Using solution \eref{eq:solution for r} and covariance \eref{eq:etaeta cov} one can see that the connected correlation of density
\begin{equation}
 \llangle \rho(x,\tau)\rho(y,0)\rrangle_c=\llangle r(x,\tau)r(y,0)\rrangle=\frac{1}{L}c(x,y,\tau)
\label{eq:rhorho hydro}
\end{equation}
is given by, for $\tau\ge 0$,
\begin{equation}
c(x,y,\tau)=\int_{0}^{\infty}\!\!\!\!ds\!\!\int_0^1 \!\!dz\,\sigma(\overline{\rho}(z))\left[\partial_{ z}G(x,z,\tau+s)\right]\left[\partial_{z}G(y,z,s)\right]
\label{eq:c}
\end{equation}

The correlation of the current can be obtained in a similar way. The average current $\llangle J \rrangle= -D(\overline{\rho}(x))\partial_x\overline{\rho}(x)$ and, because the system is one dimensional, it is constant (independent of $x$) in the steady state. Considering small fluctuations in \eref{eq:fhd 0} one writes a deviation of the current from its average value
\begin{equation}
\delta J(x,\tau)\simeq \eta(x,\tau)-\partial_x\left[D(\overline{\rho}(x))r(x,\tau)\right].
\label{eq:dJ}
\end{equation}
The connected correlation of the current is defined as $\llangle J(x,\tau)J(y,\tau^{\prime}) \rrangle_c=\llangle \delta J(x,\tau)\delta J(y,\tau^{\prime}) \rrangle$. Keeping up to the leading order term in small fluctuations one gets
\begin{eqnarray}
 \fl  \left \langle \!J(x,\!\tau)J(y,\!\tau^{\prime}) \!\right\rangle_c\!\!\!=\!\!\frac{1}{L}\!\bigg[\!\sigma(\overline{\rho}(y))\delta(x\!\!-\!\!y)\delta(\tau\!\!-\!\!\tau^{\prime})
\!\!+\!\!f(x,y,\tau\!\!-\!\!\tau^{\prime}) \Theta(\tau\!\!-\!\!\tau^{\prime})\!\!+\!\!f(y,x,\tau^{\prime}\!\!-\!\!\tau) \Theta(\tau^{\prime}\!\!-\!\!\tau)\!\bigg]
 \label{eq:curr corr fhd 2}
\end{eqnarray}
where the delta function comes from the covariance \eref{eq:etaeta cov} and $f(x,y,\tau)$ is
\begin{eqnarray}
\fl f(x,y,\tau)= -\sigma(\overline{\rho}(y))\partial_x\bigg[D(\overline{\rho}(x))\partial_{y}G(x,y,\tau)\bigg]+\partial_{x}\partial_{y}\bigg[D(\overline{\rho}(x))D(\overline{\rho}(y))c(x,y,\tau)\bigg]
\label{eq:f macro}
\end{eqnarray}
for $\tau\ge 0$, where we used \eref{eq:solution for r}, $\llangle \eta(x,\tau)r(y,0)\rrangle=0$ due to causality, and in writing the last term we used \eref{eq:rhorho hydro}.

\subsection{Two-time correlations in terms of equal-time correlation of density}
The equal-time correlation $c(x,y,0)=c(x,y)$ in the steady state can be obtained by taking $\tau=0$ in the general solution \eref{eq:c}. It satisfies a differential equation
\begin{equation}
\fl \qquad \qquad \partial_x^2\left[D(\overline{\rho}(x))c(x,y)\right]+\partial_y^2\left[D(\overline{\rho}(y))c(x,y)\right]=\partial_x\left[\sigma(\overline{\rho}(x))\delta^{\prime}(x-y)\right]
\label{eq:c diff}
\end{equation}
One can see this by substituting the formula of $c(x,y)$ from \eref{eq:c} and using equation \eref{eq:G} which leads to
\begin{equation*}
\fl \partial_x^2\left[D(\overline{\rho}(x))c(x,y)\right]+\partial_y^2\left[D(\overline{\rho}(y))c(x,y)\right]\!\!=\!\!\int_0^1\!\!\!\!dz\sigma(\overline{\rho}(z))\!\!\int_{0}^{\infty}\!\!\!\!\!\!ds\frac{d}{ds}\!\left\{\left[\partial_zG(x,z,s)\right]\!\left[\partial_zG(y,z,s)\right]\right\}
\end{equation*}
The integration over $s$ is done using the boundary values $G(x,z,\infty)=0$ and $G(x,z,0)=\delta(x-z)$. Subsequently, using an integration by parts one obtains the equation \eref{eq:c diff}.

The correlation $c(x,y)$ is often written \cite{Spohn1983,Eyink1990,Bertini2009,Jona-Lasinio2010} as a sum of an equilibrium part and an out-of-equilibrium part. This can be done using \eref{eq:c} for $\tau=0$, an integration by parts, and subsequently \eref{eq:G2} which leads to
\begin{eqnarray*}
\fl 2\,c(x,y)=c(x,y)+c(y,x)=-\int_{0}^{1}dz\int_0^{\infty}ds\frac{\sigma(\overline{\rho}(z))}{D(\overline{\rho}(z))}\frac{d}{ds}\left[G(x,z,s)G(y,z,s)\right]\cr
\qquad \qquad \qquad \qquad -\int_{0}^{1}dz\int_0^{\infty}ds\left[\frac{d}{dz}\sigma(\overline{\rho}(z))\right]\frac{d}{dz}\left[G(x,z,s)G(y,z,s)\right].
\end{eqnarray*}
In the first term on the right hand side the integration over $s$ is simple and the result is expressed in terms of the Green's function at two limits of the integration. Only the contribution from lower-limit is non-zero which one writes using $G(u,v,0)=\delta(u-v)$. In the second term one can use $\partial_z\sigma(\overline{\rho}(z))=\sigma^{\prime}(\overline{\rho}(z))\partial_z\overline{\rho}(z)$ and replace $\partial_z\overline{\rho}(z)$ using the average current $\llangle J\rrangle=-D(\overline{\rho}(z))\partial_z\overline{\rho}(z)$ which leads to a formula
\begin{equation}
c(x,y)=\frac{\sigma(\overline{\rho}(y))}{2D(\overline{\rho}(y))}\delta(x-y)+B(x,y)
\label{eq:c B}
\end{equation}
with
\begin{equation}
B(x,y)=-\llangle J \rrangle\int_0^{1}dz\frac{d}{dz}\left[\frac{\sigma^{\prime}(\overline{\rho}(z))}{2D(\overline{\rho}(z))}\right]\int_0^{\infty}dsG(x,z,s)G(y,z,s)
\label{eq:B}
\end{equation}
where in the last formula we used an integration by parts. (Generalizing the fluctuating hydrodynamics approach in higher dimensions or in presence of an external field one can write a similar formula for the correlation.)

At equilibrium, the long-range part $B(x,y)$ of the correlation vanishes because $\llangle J \rrangle=0$. (One may note that for some diffusive systems like random walk or zero range process, where $\sigma^{\prime}(\rho)=2D(\rho)$ \cite{Bodineau2004}, $B(x,y)$ vanishes even out of equilibrium.) One can show using \eref{eq:G} that it satisfies a differential equation
\begin{equation*}
\partial_x^2\left[D(\overline{\rho}(x))B(x,y)\right]+\partial_y^2\left[D(\overline{\rho}(y))B(x,y)\right]=\delta(x-y)\llangle J \rrangle\frac{d}{dx}\left[\frac{\sigma^{\prime}(\overline{\rho}(x))}{2D(\overline{\rho}(x))}\right]
\label{eq:B diff}
\end{equation*}
This is a special case of a general result \cite{Bertini2009,Jona-Lasinio2010} in an arbitrary dimension obtained using the macroscopic fluctuation theory. The long-range nature of $B(x,y)$ can be easily seen in a system with constant diffusivity where one may think of $B(x,y)$ as the electrostatic potential due to localized charges along the diagonal $y=x$.  

It is often useful to rewrite the two-time correlations \eref{eq:c}-\eref{eq:f macro} in terms of the equal-time correlation $c(x,y)$. From the identity \eref{eq:id G one}
one can rewrite \eref{eq:c} as
\begin{equation}
c(x,y,\tau)=\int_{0}^{1}dz G(x,z,\tau)c(z,y)
\label{eq:c inter}
\end{equation}
Similarly, formula \eref{eq:f macro} leads to
\begin{equation}
f(x,y,\tau)=\int_0^1 dz\frac{d}{dx}\left[D(\overline{\rho}(x))G(x,z,\tau)\right]A(z,y)
\label{eq:f inter}
\end{equation}
where we defined
\begin{equation}
A(z,y)=-\sigma(\overline{\rho}(y))\delta^{\prime}(y-z)+\frac{d}{dy}\left[D(\overline{\rho}(y))c(z,y)\right]
\label{eq:Axy}
\end{equation}

\subsection{Long and short time limits \label{sec:LST}}

At long time $\tau\gg 1$, the Green's function decays exponentially with time with a rate given by the largest eigenvalue of the operator \eref{eq:G}.
As a result, both $c(x,y,\tau)$ and $f(x,y,\tau)$ decay exponentially at large time.

At short time $\tau \ll 1$, the Green's function takes a form
\begin{equation*}
\fl G(y+\xi,y,\tau)= \frac{e^{-\frac{\xi^2}{4\tau D(\overline{\rho}(y))}}}{\sqrt{4\pi  D(\overline{\rho}(y))}}\left\{\frac{1}{\sqrt{\tau}}+\left[\frac{\left(\frac{\xi}{\sqrt{\tau}}\right)^3}{8 D(\overline{\rho}(y))}-\frac{1}{4}\left(\frac{\xi}{\sqrt{\tau}}\right) \right]\frac{\partial_yD(\overline{\rho}(y))}{D(\overline{\rho}(y))}+\Or(\sqrt{\tau}) \right\}
\end{equation*}
The curly bracket is dominated by the first term $\frac{1}{\sqrt{\tau}}$, in the range $\xi \sim \sqrt{\tau}$. (Outside this range the Green's function is essentially zero due to the exponential prefactor.) So a legitimate approximation of the Green's function at short time is
\begin{equation}
 G(x,y,\tau)\simeq \frac{e^{-\frac{(x-y)^2}{4\tau D(\overline{\rho}(y))}}}{\sqrt{4\pi \tau  D(\overline{\rho}(y))}} \qquad \textrm{at $\tau\ll 1$}
\label{eq:G asymp}
\end{equation}
Substituting this into \eref{eq:c B} and \eref{eq:c inter} one gets
\begin{equation}
c(x,y,\tau)\simeq \frac{\sigma(\overline{\rho}(y))}{2D(\overline{\rho}(y))}\frac{e^{-\frac{(x-y)^2}{4\tau D(\overline{\rho}(y))}}}{\sqrt{4\pi \tau  D(\overline{\rho}(y))}}+B(x,y) \qquad \textrm{for $\tau\ll 1$}
\label{eq:c final final}
\end{equation}
where the last term is obtained by using that $B(x,y)$ varies slowly with $x$.
In the limit $\tau\rightarrow 0$ one gets back the stationary correlation \eref{eq:c B}.

To calculate the asymptotics of $f(x,y,\tau)$ one substitutes \eref{eq:c final final} in \eref{eq:f macro}, and rearranging the terms one gets
\begin{eqnarray*}
\fl f(x,y,\tau)\simeq -\frac{\sigma(\overline{\rho}(y))}{2}\frac{d^2}{dxdy}\left[D(\overline{\rho}(x))G(x,y,\tau)\right]+\frac{1}{2}\frac{d\sigma(\overline{\rho}(y))}{dy}\frac{d}{dx}\left[D(\overline{\rho}(x))G(x,y,\tau)\right]\cr
\qquad \qquad \qquad \qquad \qquad  +\frac{d^2}{dxdy}\left[ D(\overline{\rho}(x)) D(\overline{\rho}(y))B(x,y)\right]
\end{eqnarray*}
For small $\tau$, using \eref{eq:G asymp} one can see $\partial_x\partial_y\left[D(\overline{\rho}(x)) G(x,y,\tau)\right]\sim \tau^{-\frac{3}{2}}$ and $\partial_x \left[D(\overline{\rho}(x)) G(x,y,\tau)\right]\sim \tau^{-1}$ in the range $x-y\sim \sqrt{\tau}$. This gives
\begin{equation}
\fl  f(x,y,\tau)\simeq \frac{1}{\tau^{\frac{3}{2}}}\frac{e^{-\frac{(x-y)^2}{4D(\overline{\rho}(y))\tau}}}{\sqrt{4\pi D(\overline{\rho}(y))}}\left[\frac{(x-y)^2}{4D(\overline{\rho}(y))\tau}-\frac{1}{2}\right]\frac{\sigma(\overline{\rho}(y))}{2}+\frac{d^2\left[ D(\overline{\rho}(x))D(\overline{\rho}(y))B(x,y)\right]}{dxdy}
\label{eq:f short}
\end{equation}
Outside the range $x-y\sim \sqrt{\tau}$ the first term on the right hand side is essentially zero, and the leading contribution comes from the last term.

\subsection{Integrated current}
The integrated current in the hydrodynamic scale is defined as
\begin{equation}
q(x,\tau)=\int_{0}^{\tau}dsJ(x,s)
\label{eq:qJ}
\end{equation} 
and so its connected correlation is
\begin{equation*}
\llangle q(x,\tau)q(y,\tau^{\prime})\rrangle_c=\int_{0}^{\tau}ds_1\int_{0}^{\tau^{\prime}}ds_2 \llangle J(x,s_1)J(y,s_2) \right\rangle_c.
\end{equation*}
Using \eref{eq:curr corr fhd 2} and considering $\tau\ge \tau^{\prime}$ one can perform the integration, leading to
\begin{eqnarray}
\fl \llangle q(x,\tau)q(y,\tau^{\prime})\rrangle_c=\frac{1}{L}\Bigg[\sigma(\overline{\rho}(y))\delta(x-y)\tau^{\prime}+\int_0^{\tau^{\prime}}ds(\tau^{\prime}-s)f(y,x,s)\cr \qquad \qquad \qquad +\tau^{\prime}\int_0^{\tau-\tau^{\prime}}ds f(x,y,s)+\int_{\tau-\tau^{\prime}}^{\tau}ds (\tau-s)f(x,y,s)\Bigg]
\label{eq:qq hydro 0}
\end{eqnarray}

\subsubsection*{Long and short time limits:}
To analyze the asymptotics it is convenient to rewrite \eref{eq:qq hydro 0} in an alternative form. This is done using an identity
 \begin{equation}
\sigma(\overline{\rho}(x))\delta(x-y)=\int_{0}^{1}dz\sigma(\overline{\rho}(z))-\int_0^{\infty}ds\left[f(y,x,s)+f(x,y,s)\right]
\label{eq:f+f id}
\end{equation}
proved in the \ref{sec:f+f id}. Following a straightforward algebra one writes
\begin{eqnarray}
\fl \llangle q(x,\tau)q(y,\tau^{\prime})\rrangle_c=\frac{\tau^{\prime}}{L}\int_{0}^{1}dz\sigma(\overline{\rho}(z)) +\frac{1}{L}\Bigg\{-\int_{0}^{\infty}ds\min(\tau^{\prime},s)f(y,x,s)\cr
\qquad \quad -\int_{0}^{\infty}ds\min(\tau,s)f(x,y,s)+\int_{0}^{\infty}ds\min(\tau-\tau^{\prime},s)f(x,y,s)\Bigg\}
\label{eq:qq hydro}
\end{eqnarray}
for $\tau\ge \tau^{\prime}$.
Under this form it is easy to analyze the contribution at large and at short times, as discussed below. 

At large times $\tau\ge \tau^{\prime}\gg 1$, the leading contribution comes from the first term on the right hand side of \eref{eq:qq hydro} which grows linearly with time, whereas the remaining terms are bounded. One may see this using
\begin{equation*}
\lim_{t\rightarrow\infty}\int_0^{\infty}ds\min(t,s)f(u,v,s)=\int_{0}^{\infty}ds\; s \; f(u,v,s)+\lim_{t\rightarrow\infty}t\int_{t}^{\infty}dsf(u,v,s)
\end{equation*}
which remains finite as $f(u,v,s)$ decays exponentially for large $s$ (see \sref{sec:LST}). Thereafter
\begin{equation}
\llangle q(x,\tau)q(y,\tau^{\prime})\rrangle_c\simeq\frac{\tau^{\prime}}{L}\int_{0}^{1}dz\sigma(\overline{\rho}(z)) \qquad \textrm{for $\tau\ge \tau^{\prime}\gg 1$}.
\label{eq:qq long}
\end{equation}
This agrees with an earlier result \cite{Bodineau2004} for the variance of integrated current. Note that at large times $\tau$ and $\tau^{\prime}$, the leading behavior does not depend on $x$ and $y$.

\begin{figure}
\begin{center}
\includegraphics[scale=0.6]{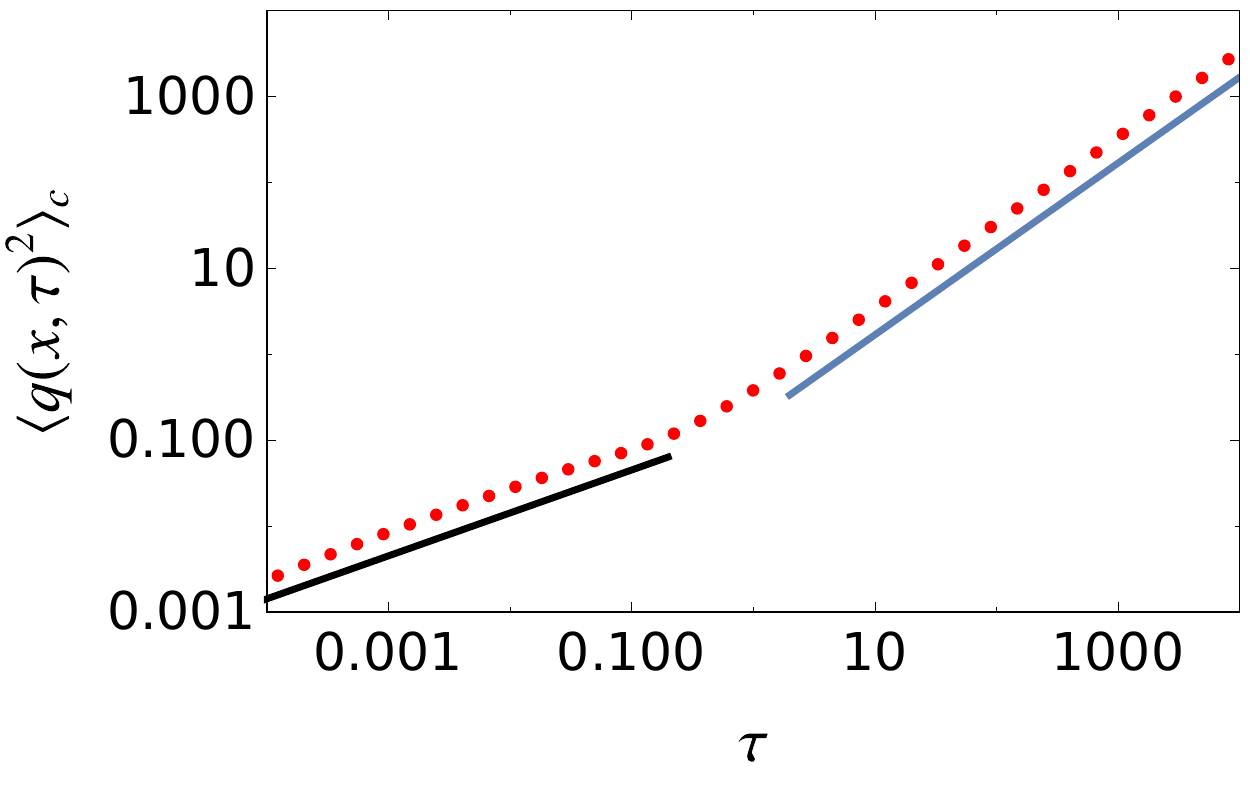}
\end{center}
\caption{A numerical plot of $\llangle q(x,\tau)^2\rrangle_c$ in \eref{eq:qq hydro} for the symmetric exclusion process at $x=\frac{1}{2}$ and $(\rho_a,\rho_b)=(1,0)$. The straight lines show the asymptotic behavior in \eref{eq:qq long} and \eref{eq:q22}. \label{fig:f short}}
\end{figure}

The behavior at short time is also interesting, particularly, because, it reproduces the behavior in an infinite system. It is known that, in an infinite system at equilibrium, the variance $\llangle q(x,\tau)^2\rrangle_c$ scales sub-diffusively as $\sqrt{\tau}$ \cite{Gerschenfeld2009Bethe,Gerschenfeld2009}. One finds a similar dependence at short time in \eref{eq:qq hydro}. It is simpler to analyze the correlation at the same site $y=x$. The leading contribution at short time comes from the term inside the curly brackets. One may see this by writing
\begin{equation*}
\fl \int_0^{\infty}ds\min(\tau,s)f(x,x,s)=\int_0^{\tau}ds\,s\,f(x,x,s)+\tau\int_{\tau}^{1}ds f(x,x,s)+\tau\int_1^{\infty}dsf(x,x,s)
\end{equation*}
The last integral is finite; thus the contribution is sub-dominant compared to the leading $\sqrt{\tau}$ dependence of the first two terms. At short $s$, \eref{eq:f short} gives
\begin{equation*}
f(x,x,s)\simeq -\frac{1}{s^{\frac{3}{2}}}\frac{\sigma(\overline{\rho}(x))}{8\sqrt{\pi D(\overline{\rho}(x)))}}
\end{equation*}
and substituting this in the first two terms one gets
\begin{equation*}
 \int_{0}^{\infty}ds\min(\tau,s)f(x,x,s)= -\frac{\sigma(\overline{\rho}(x))}{2\sqrt{\pi D(\overline{\rho}(x))}}\sqrt{\tau}+\Or(\tau) \qquad \textrm{at $\tau\ll 1$.}
\end{equation*}
Substituting in \eref{eq:qq hydro} one gets the correlation at short time
\begin{equation}
 \llangle q(x,\tau)q(x,\tau^{\prime}) \rrangle_c\simeq \frac{1}{L}\left[ \sqrt{\tau} +\sqrt{\tau^{\prime}} -\sqrt{\vert\tau-\tau^{\prime}\vert} \right]\frac{\sigma(\overline{\rho}(x))}{2\sqrt{\pi\,D(\overline{\rho}(x))}}.
\label{eq:q22}
\end{equation}
The formula is in agreement with an earlier result \cite{Gerschenfeld2009,Gerschenfeld2009Bethe} for the second cumulant of the integrated current in an infinite line. A comparison of the asymptotics \eref{eq:qq long} and \eref{eq:q22} with a numerical plot of the whole function \eref{eq:qq hydro} is given in \fref{fig:f short} in the case of the symmetric exclusion process.

\subsection{The symmetric exclusion process using the fluctuating hydrodynamics\label{sec:sep explicit}}
Fluctuating hydrodynamics for the symmetric exclusion process corresponds to $D(\rho)=1$ and $\sigma(\rho)=2\rho(1-\rho)$ \cite{Bodineau2004,Derrida2007}. The average density profile and the average current in the steady state
\begin{equation}
\overline{\rho}(x)=\rho_a-(\rho_a-\rho_b)x \qquad \textrm{and}\qquad \llangle J \rrangle=(\rho_a-\rho_b)
\label{eq:ave density hydro}
\end{equation}
respectively, which one obtains as a solution of \eref{eq:av profile}.

Because $D(\rho)=1$, one can get an explicit expression of the Green's function by solving \eref{eq:G}
\begin{equation}
G(x,y,\tau)=2\sum_{n=1}^{\infty}e^{-n^2\pi^2\tau}\sin(n\pi x)\sin(n \pi y)\qquad \textrm{for all $\tau\ge 0$}.
\label{eq:G explicit}
\end{equation}
Moreover, substituting the corresponding $D(\rho)$ and $\sigma(\rho)$ for the symmetric exclusion process in \eref{eq:B} one gets
\begin{equation}
\fl  B(x,y)=-2(\rho_a-\rho_b)^2\int_0^{\infty}\!\!\!ds\int_0^1\!\!\!dzG(x,z,s)G(y,z,s)=-(\rho_a-\rho_b)^2\int_{0}^{\infty}\!\!\!ds  G(x,y,s)
\label{eq:B explicit}
\end{equation}
where in the last equality we used \eref{eq:id G one}. Substituting results (\ref{eq:G explicit},\ref{eq:B explicit}) in \eref{eq:c B} one gets (Alternatively, one may use the formulas (\ref{eq:U},\ref{eq:U explicit}).)
\begin{equation}
\fl c(x,y)=\overline{\rho}(y)(1-\overline{\rho}(y))\delta(x-y)-(\rho_a-\rho_b)^2\left[ x(1-y)\Theta(y-x)+y(1-x)\Theta(x-y) \right]
\label{eq:c eq hydro explicit}
\end{equation}
in agreement with expressions previously derived \cite{Spohn1983,Derrida20072} using microscopic dynamics.

Substituting (\ref{eq:G explicit},\ref{eq:c eq hydro explicit}) in \eref{eq:c inter} one gets an explicit formula for the two-time correlation
\begin{equation}
\fl \qquad c(x,y,\tau)= 2\sum_{n=1}^{L}e^{-n^2\pi^2\,\tau}\sin\! \left(n\pi x\right)\!\sin\!\left(n\pi y\right)\left[\overline{\rho}(y)(1-\overline{\rho}(y))-\frac{(\rho_a-\rho_b)^2}{(n\pi)^2}\right]
\label{eq:tautau macro 0}
\end{equation}

In a similar way formulas (\ref{eq:f inter},\ref{eq:Axy}) lead to
\begin{eqnarray}
\fl \quad f(x,y,\tau)= 2\sum_{n=1}^{\infty}e^{-n^2\pi^2\tau} \bigg[-\overline{\rho}(y)(1-\overline{\rho}(y))\!\left( n\pi\right)^2\!\cos\! \left(n\pi x\right)\!\cos\!\left(n\pi y\right)\cr
\fl   \qquad  \quad  -\!\left(\rho_a\!-\!\rho_b\right)\!\left(1\!-\!2\overline{\rho}(y) \right)\!\left(n\pi\right)\!\cos\! \left(n\pi x\right)\sin\!\left(n\pi y\right)\!-\!\left(\rho_a\!-\!\rho_b\right)^2\cos\! \left(n\pi x\right)\cos\!\left(n\pi y\right)\bigg]
\label{eq:Fxy}
\end{eqnarray}
where $\overline{\rho}(z)$ is given in \eref{eq:ave density hydro}.

\subsection{Spectral distribution}
One characterization of the fluctuations of the current across a bond is the spectral power density
\begin{equation}
S(x,\nu)=\lim_{T\rightarrow \infty}\frac{1}{T}\left[\llangle \vert \widehat{J}(x,\nu) \vert^2 \rrangle- \vert\llangle \widehat{J}(x,\nu) \rrangle\vert^2\right]
\label{eq:spectral power}
\end{equation}
where $\widehat{J}(x,\nu)$ is the Fourier transform of $J(x,\tau)$ in a time window $[0,T]$.
The Wiener-Khintchine theorem \cite{Kubo2} asserts that
\begin{equation}
S(x,\nu)=\int_{-\infty}^{\infty}d\tau\,e^{-\hat{i}2\pi\nu \tau}\llangle J(x,\tau) J(x,0) \rrangle_c
\label{eq:spectral density one}
\end{equation}

\begin{figure}
\begin{center}
\includegraphics[width=0.6\linewidth]{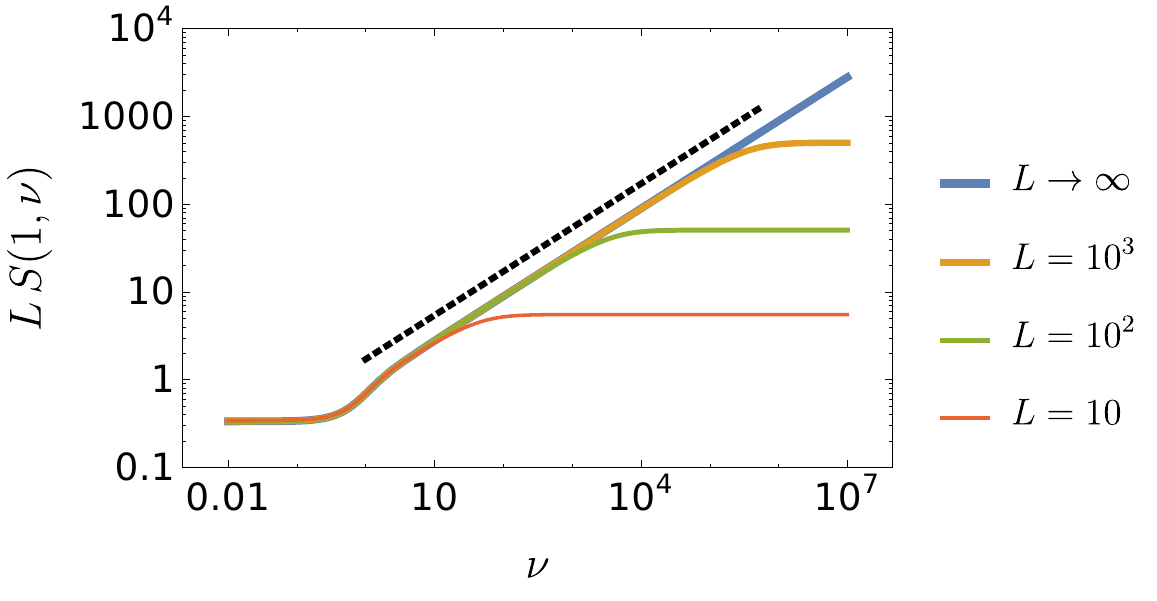}
\end{center}
\caption{A numerical plot of $L\,S(1,\nu)$ in \eref{eq:S0 exact} for $(\rho_a,\rho_b)=(0,\frac{1}{2})$ and for a set of systems of length $L$ indicated in the figure. The graph for $L\rightarrow \infty$ is the function in \eref{eq:S0}. The dotted straight line denotes a power-law $\sqrt{\nu}$. \label{fig:power spectra}}
\end{figure}

To begin, one uses \eref{eq:curr corr fhd 2}, and writes
\begin{eqnarray*}
\fl \int_{-\infty}^{\infty}d\tau\,e^{-\hat{i}2\pi\nu \tau}\llangle J(x,\tau) J(y,0) \rrangle_c\cr
\fl \qquad \qquad =\frac{1}{L}\left\{\sigma(\overline{\rho}(y))\delta(x-y)+\int_0^{\infty}d\tau e^{-\hat{i}2 \pi \nu \tau}f(x,y,\tau)+\int_0^{\infty}d\tau e^{\hat{i}2 \pi \nu \tau}f(y,x,\tau)\right\}
\end{eqnarray*} 
Using \eref{eq:f+f id} and in the limit $y\rightarrow x$ one gets
\begin{equation}
 S(x,\nu)=\frac{1}{L}\left\{\int_0^{1}dz\sigma(\overline{\rho}(z))+2\int_0^{\infty}d\tau \left[\cos (2\pi \nu \tau)-1\right] f(x,x,\tau)\right\}
\label{eq:Sxn}
\end{equation}
where $f(x,x,s)$ is given in \eref{eq:f inter}.

To obtain a more explicit expression one can use the formula \eref{eq:Fxy} in \eref{eq:Sxn} for the symmetric exclusion process. For example, at the right boundary $x=1$ one gets
\begin{eqnarray*}
\fl  S(1,\nu)\!\!=\!\!\frac{1}{L}\left\{\!\int_0^{1}\!\!dz\sigma(\overline{\rho}(z))+ \!\!\sum_{n=1}^{\infty}\frac{\rho_b(1-\rho_b)16\pi^2 \nu^2}{(n\pi)^4+4\pi^2\nu^2}\!+\!(\rho_a-\rho_b)^2\!\left[\frac{2}{3}-\!\!\sum_{n=1}^{\infty}\frac{4\pi^2n^2}{(n\pi)^4+4\pi^2\nu^2}  \right]\!\right\}
\end{eqnarray*}
For the symmetric exclusion process one has
\begin{equation*}
\int_0^{1}dz\sigma(\overline{\rho}(z))=\rho_a+\rho_b-2\rho_a\rho_b-\frac{2}{3}(\rho_a-\rho_b)^2
\end{equation*}
Substituting this one gets
\begin{eqnarray}
  S(1,\nu)=\frac{1}{L}\left\{\rho_a+\rho_b-2\rho_a\rho_b+\rho_b(1-\rho_b) \sum_{n=1}^{\infty}\frac{16\pi^2 \nu^2}{(n\pi)^4+4\pi^2\nu^2}\right. \cr \left.\qquad \qquad\qquad\qquad \qquad \qquad \quad -(\rho_a-\rho_b)^2\sum_{n=1}^{\infty}\frac{4\pi^2n^2}{(n\pi)^4+4\pi^2\nu^2}  \right\}
\label{eq:S0}
\end{eqnarray}

The first sum in \eref{eq:S0} diverges as $\sqrt{\nu}$ when $\nu\rightarrow \infty$. This is because $\llangle J(x,0)J(x,0)\rrangle$ is ill defined (due to the Dirac delta functions $\delta(x-y)\delta(\tau-\tau^{\prime})$ in \eref{eq:curr corr fhd 2}). One must realize that \eref{eq:S0} comes from a hydrodynamic theory which is only defined above a certain coarse grained scale. This sets an upper cutoff in the summation of the eigenmodes $n$, which leads to saturation of $S(1,\nu)$ at large $\nu$. One can see this in \fref{fig:power spectra} where we compare \eref{eq:S0} with an exact formula of $S(1,\nu)$ derived in \ref{eq:exact derivation} using the microscopic details of the symmetric exclusion process,
\begin{eqnarray}
\fl S(1,\nu)\simeq \frac{1}{L}\left\{\rho_a+\rho_b-2\rho_a\rho_b+\rho_b(1-\rho_b) \sum_{n=1}^{L}\frac{4\pi^2 \nu^2\,\left(\cos\frac{\theta_n}{2}\right)^2}{4L^4\left(\sin\frac{\theta_n}{2}\right)^4+\pi^2\nu^2}\right. \cr \left.\qquad \qquad\qquad\qquad\qquad  -(\rho_a-\rho_b)^2\sum_{n=1}^{L}\frac{L^2\left(\sin\theta_n\right)^2}{4L^4\left(\sin\frac{\theta_n}{2}\right)^4+\pi^2\nu^2}  \right\}
\label{eq:S0 exact}
\end{eqnarray}
where $\theta_n=\frac{n\pi}{L+1}$. At large $\nu$, formula \eref{eq:S0 exact} saturates, as shown in the numerical plot in \fref{fig:power spectra}.

\section{The symmetric exclusion process using microscopic dynamics \label{sec:sep}}
\begin{figure}
\begin{center}
\includegraphics[width=0.8\textwidth]{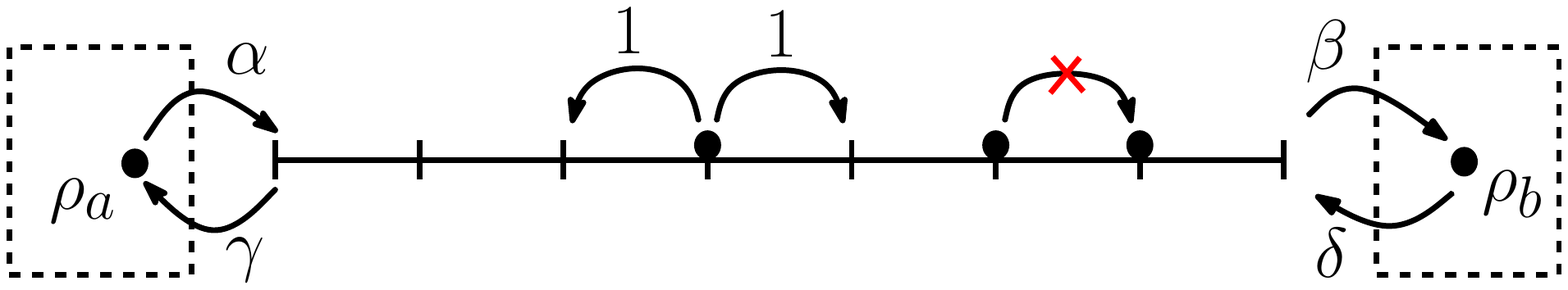}
\end{center}
\caption{Jump rates in the symmetric exlusion process coupled with a reservoir of density $\rho_a$ at the left boundary and a reservoir of density $\rho_b$ at the right boundary. \label{fig:sep}}
\end{figure}

The symmetric exclusion process is defined on a one-dimensional lattice of $L$ sites, coupled with two reservoirs of density $\rho_a$ and $\rho_b$. Particles within the bulk jump from a site to one of its neighbor sites as long as the jump respects simple exclusion: at any time there could be at most one particle at a site. The time scale is set by choosing bulk jump rates equal to $1$. (This is equivalent to setting $D(\rho)=1$ in the hydrodynamic limit.) At the boundary sites, the coupling to the reservoirs is modeled by injection and extraction rates of particles as shown in \fref{fig:sep}; the density of the reservoirs are related \cite{Derrida2007} to these boundary rates by
\begin{equation*}
\rho_a=\frac{\alpha}{\alpha+\gamma}; \qquad \rho_b=\frac{\delta}{\beta+\delta}
\end{equation*}

At long time, the system reaches a steady state where the average occupation per site is linear in space, and the average current is constant \cite{Spohn1983,Derrida2007,Derrida20072},
\begin{equation}
\fl \qquad \llangle \tau_i \rrangle=\frac{1}{N}\left[\rho_a\left(L+b-i\right)+\rho_b\left(i-1+a\right)\right]; \qquad \llangle \mathcal{J}_i\rrangle\equiv \llangle \mathcal{J}\rrangle=\frac{(\rho_a-\rho_b)}{N}
\label{eq:average results}
\end{equation}
where we defined 
\begin{equation}
a=\frac{1}{\alpha+\gamma};\qquad b=\frac{1}{\beta+\delta};\qquad \textrm{and}\quad N=L+a+b-1
\label{eq:ab} 
\end{equation}

The steady state two-point correlation function of occupation variables as well as higher order correlation functions are known \cite{Spohn1983,Derrida2007,Derrida20072}. For example,
\begin{equation}
\llangle \tau_{i}\tau_{j}\rrangle_c=\cases{-\left[\frac{(\rho_a-\rho_b)^2}{N^2 (N-1)}\right](i+a-1)(L+b-j) & for $i<j$\cr
\llangle \tau_i \rrangle(1-\llangle \tau_i \rrangle) & for $i=j$}
\label{eq:stationary tautau}
\end{equation}
which reduces to \eref{eq:c eq hydro explicit} in the large $L$ limit.

Using the definition of the model, one can write the time evolution of these averages. For example,
\begin{equation}
\frac{d\llangle\boldsymbol{\tau}(t)\rrangle}{dt}=-\boldsymbol{M}\cdot \llangle\boldsymbol{\tau}(t)\rrangle+\mathbf{B}
\label{eq:eta evolution}
\end{equation}
where
\begin{equation}
\fl  \boldsymbol{\tau}(t)=\left( \begin{array}{c} \tau_1(t) \\ \tau_2(t) \\ . \\ . \\ . \\ \tau_L(t) \end{array} \right); \; \mathbf{M}=\left( \begin{array}{ccccccc} 1+\frac{1}{a} & -1 & 0 & . & . & . & . \\ -1 & 2 & -1 & 0 & . & . & .\\0 & -1 & 2 & -1 & . & . & .\\ . & . & . & . & . & . & . \\ . & . & . & 0 & -1 & 2 & -1\\ . & . & . & . & 0 & -1 & 1+\frac{1}{b} \end{array} \right);\; \mathbf{B}=\left( \begin{array}{c} \alpha \\ 0 \\ . \\ 0 \\ \delta \end{array} \right)
\label{eq:matrix M explicit}
\end{equation}
We shall use bold font to denote matrices.

The solution of \eref{eq:eta evolution} is simple to write in terms of a microscopic Green's function
$\mathbf{g}(t)=\left[g_{i,j}(t)\right]_{L\times L}$
defined as the solution of
\begin{equation}
\frac{d\mathbf{g}(t)}{dt}+\boldsymbol{M}\cdot \mathbf{g}(t)=\delta(t)\mathbf{1}
\label{eq:Green's function}
\end{equation}
where $\mathbf{1}$ is the identity matrix. The solution of \eref{eq:eta evolution} is
\begin{equation}
\llangle \boldsymbol{\tau}(t)\rrangle=\mathbf{g}(t)\cdot\llangle \boldsymbol{\tau}(0)\rrangle+\int_{0}^{t}dt^{\prime}\mathbf{g}(t-t^{\prime})\cdot \mathbf{B}
\label{eq:tau sol}
\end{equation}

The integrated current $Q_i(t)$ is defined as the number of particles passed across the bond $(i,i+1)$ during a time window $[0,t]$. Its average $\llangle Q_i(t)\rrangle$ satisfies
\begin{equation}
\frac{d\llangle Q_i(t) \rrangle}{dt}=
\llangle \tau_i(t)\rrangle -\llangle \tau_{i+1}(t)\rrangle \qquad \textrm{ for $1\le i \le L-1$.}
\label{eq:Q evolution}
\end{equation}

\subsection{Correlation of the occupation variable}
The evolution of the two-time correlation $\llangle\tau_i(t)\tau_j(s)\rrangle_c$ of the occupation variables is given by
\begin{equation*}
\frac{d\mathbf{C}(t,s)}{dt}=-\boldsymbol{M}\cdot \mathbf{C}(t,s)\qquad \textrm{for $t\ge  s$}
\end{equation*}
where $\mathbf{C}(t,s)\equiv \left[\llangle \tau_{i}(t)\tau_{j}(s) \rrangle_c\right]_{L\times L}$.
The solution can be written in terms of the microscopic Green's function,
\begin{equation}
\llangle \tau_{i}(t)\tau_{j}(s) \rrangle_c=\sum_{\ell=1}^{L}g_{i,\ell}(t-s) \llangle \tau_{\ell}(s)\tau_{j}(s) \rrangle_c .
\label{eq:tt correlation}
\end{equation}
This way the two-time correlation is expressed in terms of the known \cite{Spohn1983,Derrida20072,Derrida2007} equal-time correlations. One may compare \eref{eq:tt correlation} with the hydrodynamic formula \eref{eq:c inter}.

It is interesting to write the microscopic equivalent of the hydrodynamic formula \eref{eq:c diff} for the steady state correlations. Starting with the jump rates one gets in the steady state 
\begin{equation}
\fl \qquad \left[\llangle \tau_{i+1}\tau_j\rrangle_c-2\llangle \tau_{i}\tau_j\rrangle_c+\llangle \tau_{i-1}\tau_j\rrangle_c\right]+\left[\llangle \tau_{i}\tau_{j+1}\rrangle_c-2\llangle \tau_{i}\tau_j\rrangle_c+\llangle \tau_{i}\tau_{j-1}\rrangle_c\right]=\Omega_{i,j}
\label{eq:corr diff micro}
\end{equation}
for bulk sites, where 
\begin{equation}
\fl \Omega_{i,j}=\llangle (\tau_i-\tau_{i+1})^2 \rrangle\delta_{i+1,j}+\llangle (\tau_i-\tau_{i-1})^2\rrangle\delta_{i-1,j}+2\left[\llangle \tau_i\tau_{i+1}\rrangle-2\llangle\tau_i\rrangle+\llangle \tau_i\tau_{i-1}\rrangle \right]\delta_{i,j}
\end{equation}
To see how one can get \eref{eq:c diff}, one uses the stationary averages (\ref{eq:average results},\ref{eq:stationary tautau}) and gets, for large $L$,
\begin{equation*}
\fl \Omega_{i,j}\simeq 2\llangle \tau_i\rrangle (1-\llangle \tau_i\rrangle)\left[\delta_{i+1,j}-2\delta_{i,j}+\delta_{i-1,j}\right]-\frac{(\rho_a-\rho_b)}{L}(1-2\llangle \tau_i\rrangle)\left(\delta_{i+1,j}-\delta_{i-1,j}\right)
\end{equation*}
In the hydrodynamic limit
\begin{equation*}
\Omega_{i,j}\simeq \frac{1}{L^3}\left[2\overline{\rho}(x)(1-\overline{\rho}(x))\delta^{\prime\prime}(x-y)+2\frac{d\overline{\rho}(x)}{dx}(1-2\overline{\rho}(x))\delta^{\prime}(x-y)\right]
\end{equation*}
where one uses $\overline{\rho}(x)=\rho_a-(\rho_a-\rho_b)x$ for the symmetric exclusion process. Clearly, taking the hydrodynamic limit of \eref{eq:corr diff micro} and using $\llangle \tau_i\tau_j\rrangle_c\simeq \frac{1}{L}c(x,y)$ one gets \eref{eq:c diff}.

\subsection{Correlation of the integrated current}
One can write a formula similar to \eref{eq:tt correlation} to express $\llangle Q_i(t)Q_j(s)\rrangle_c$ in terms of $\llangle \tau_i(s)\tau_j(s)\rrangle_c$. We first present the result and then give a derivation in the latter part of this section. For simplicity of presentation we shall only consider current across bonds inside the bulk.

\numparts
Without loss of generality we take $t\ge s$. We shall show that
\begin{eqnarray}
\fl \llangle Q_i(t)Q_j(s)\rrangle_c\!=\!\delta_{i,j}\!\int_{0}^{s}\!\!\!\! dt^{\prime}\Gamma_{i}(t^{\prime})\!+\!\int_{0}^{s}\!\!\!\!dt^{\prime}\!\!\!\int_{0}^{t^{\prime}}\!\!\!\!dt^{\prime\prime}\left[F_{i,j}(t^{\prime},t^{\prime\prime})\!+\!F_{j,i}(t^{\prime},t^{\prime\prime})\right]
\!+\!\!\!\int_{s}^{t}\!\!\!\!dt^{\prime}\!\!\int_{0}^{s}\!\!\!\!dt^{\prime\prime}F_{i,j}(t^{\prime},t^{\prime\prime})
 \label{eq:formal solution}
\end{eqnarray}
where
\begin{equation}
 F_{i,j}(t^{\prime},t^{\prime\prime})=\sum_{\ell}\left[g_{i,\ell}(t^{\prime}-t^{\prime\prime})-g_{i+1,\ell}(t^{\prime}-t^{\prime\prime})\right]A_{\ell,j}(t^{\prime\prime}) \qquad \textrm{for $t^{\prime}\ge t^{\prime\prime}$}
\label{eq:expression F}
\end{equation}
\begin{equation}
A_{i,j}(t)=\llangle \tau_i(t)\tau_j(t) \rrangle_c-\llangle \tau_i(t)\tau_{j+1}(t)\rrangle_c+\Gamma_j(t)\left(\delta_{i,j+1}-\delta_{i,j}\right)
\label{eq:Aij}
\end{equation}
and
\begin{equation}
\Gamma_{i}(t)=\llangle \left[\tau_i(t)-\tau_{i+1}(t)\right]^2\rrangle
\label{eq:D}
\end{equation}
\endnumparts

\subsubsection*{Derivation:} To derive \eref{eq:formal solution} one starts by writing the evolution of the correlation of integrated current inside bulk.
\begin{equation}
\fl \qquad  \frac{d}{dt}\llangle Q_i(t)Q_j(s)\rrangle_c=\llangle \tau_i(t) Q_j(s)\rrangle_c-\llangle \tau_{i+1}(t) Q_j(s)\rrangle_c\equiv K_{i,j}(t,s) \qquad \textrm{for $t\ge s$}
\label{eq:Kij}
\end{equation}
where we introduced $K_{i,j}(t,s)$ for convenience of writing the formulas below. Integrating over time one obtains the solution,
\begin{equation*}
  \llangle Q_i(t)Q_j(s)\rrangle_c=\llangle Q_i(s)Q_j(s)\rrangle_c+\int_{s}^{t}dt'K_{i,j}(t',s)
\end{equation*}
In the \ref{app:derivation of A}, we shall show that the first term $\llangle Q_i(s)Q_j(s)\rrangle_c$ can also be written in terms of $K_{i,j}(t,s)$:
\begin{equation}
 \llangle Q_i(s)Q_j(s)\rrangle_c=\int_{0}^{s}dt^{\prime}\bigg\{\delta_{i,j}\Gamma_{i}(t^{\prime})+K_{i,j}(t^{\prime},t^{\prime})+K_{j,i}(t^{\prime},t^{\prime})\bigg\}
\label{eq:QiQjss}
\end{equation}
Putting together one writes
\begin{equation}
\fl   \llangle Q_i(t)Q_j(s)\rrangle_c=\delta_{i,j}\int_{0}^{s}dt^{\prime}\Gamma_{i}(t^{\prime})+\int_{0}^{s}dt^{\prime}\bigg\{K_{i,j}(t^{\prime},t^{\prime})+K_{j,i}(t^{\prime},t^{\prime})\bigg\}+\int_{s}^{t}dt'K_{i,j}(t',s)
 \label{eq:QQ first}
\end{equation}

In the next step one determines $K_{i,j}(t,s)$, equivalently, the correlation $\llangle \tau_i(t)Q_j(s)\rrangle_c$. One first writes 
\begin{equation}
\llangle \tau_i(t)Q_j(s)\rrangle_c=\sum_{\ell=1}^{L}g_{i,\ell}(t-s)\llangle \tau_{\ell}(s)Q_j(s)\rrangle_c \qquad \textrm{for $t\ge s$}
\label{eq:tautau ts}
\end{equation}
which is derived by a calculation similar to that for \eref{eq:tt correlation}.
The correlation $\llangle \tau_{\ell}(s)Q_j(s)\rrangle_c$ on the right hand side is written in terms of the microscopic Green's function.
\begin{equation}
\llangle \tau_i(s)Q_j(s)\rrangle_c=\int_{0}^{s}dt^{\prime\prime}\sum_{\ell} g_{i,\ell}(s-t^{\prime\prime})A_{\ell,j}(t^{\prime\prime})
\label{eq:tauQ solution}
\end{equation}
where $A_{i,j}(t)$ is given in \eref{eq:Aij}. This can be verified by substituting in the rate equation of $\llangle \tau_{\ell}(s)Q_j(s)\rrangle_c$ which in terms of a matrix $\widehat{\mathbf{C}}(s)\equiv [\llangle \tau_i(s)Q_j(s)\rrangle_c]_{L\times L-1}$ one writes
\begin{equation*}
\frac{d\widehat{\mathbf{C}}(s)}{ds}=-\boldsymbol{M}\cdot \widehat{\mathbf{C}}(s)+\mathbf{A}(s)
\end{equation*}
where $\mathbf{A}(s)\equiv [A_{i,j}(s)]_{L\times L-1}$ with $1\le i \le L$ and $1\le j \le L-1$.

Writing \eref{eq:tautau ts}, and \eref{eq:tauQ solution} together, and using the identity
\begin{equation}
\sum_{\ell}g_{i,\ell}(t-t')g_{\ell,j}(t'-t'')=g_{i,j}(t-t'')
\label{eq:id greens discrete}
\end{equation}
one gets
\begin{equation}
\llangle \tau_i(t)Q_j(s)\rrangle_c=\int_0^sdt^{\prime}\sum_{k}g_{i,k}(t-t^{\prime})A_{k,j}(t^{\prime})
\label{eq:tQ final}
\end{equation}
Substituting this in \eref{eq:Kij} leads to
\begin{equation}
\fl \quad K_{i,j}(t,s)=\int_0^sdt^{\prime}\sum_{\ell}\left[g_{i,\ell}(t-t^{\prime})-g_{i+1,\ell}(t-t^{\prime})\right]A_{\ell,j}(t^{\prime})=\int_0^sdt^{\prime}F_{i,j}(t,t^{\prime})
\label{eq:K F rel}
\end{equation}
where $F_{i,j}(t,s)$ is given in \eref{eq:expression F}.

In the final step one substitutes the above expression of $K_{i,j}(t,s)$ in \eref{eq:QQ first} which leads to the result \eref{eq:formal solution}.

\subsection{Correlation of the current}
The current $\mathcal{J}_i(t)$ is the time derivative of the integrated current 
\begin{equation*}
\mathcal{J}_i(t)=\frac{dQ_i(t)}{dt}
\end{equation*} 
and so the correlation function is
\begin{equation*}
\llangle \mathcal{J}_{i}(t)\mathcal{J}_{j}(s) \rrangle_c=\frac{d}{d t}\frac{d}{d s}\llangle Q_i(t)Q_j(s)\rrangle_c
\end{equation*}
As \eref{eq:formal solution} is only valid for $t\ge s$, for arbitrary $t$ and $s$ one has
\begin{equation*}
\fl \llangle Q_i(t)Q_j(s)\rrangle_c=\Theta(t-s)\left[u(s)+\int_{s}^{t}dt^{\prime} v_1(t^{\prime},s)\right]+\Theta(s-t)\left[u(t)+\int_{t}^{s}dt^{\prime} v_2(t^{\prime},t)\right]
\end{equation*}
where $\Theta(x)$ is the Heaviside theta function and where one can trivially read $u(s)$, $v_1(s)$ and $v_2(s)$ from \eref{eq:formal solution}.
Taking time derivative one gets
\begin{eqnarray*}
\fl \qquad \qquad \frac{d}{d t}\frac{d}{d s}\llangle Q_i(t)Q_j(s)\rrangle_c=\delta(t-s)\left[\frac{du(s)}{ds}-v_1(s,s)-v_2(t,t) \right]\cr
\qquad \qquad \qquad \qquad \qquad \qquad +\Theta(t-s)\frac{d}{d s} v_1(t,s)+\Theta(s-t)\frac{d}{d t} v_2(s,t)
\end{eqnarray*}
This leads to
\begin{equation}
\fl \qquad \llangle \mathcal{J}_{i}(t)\mathcal{J}_{j}(s) \rrangle_c=\delta(t-s)\delta_{i,j}\Gamma_i(s)+\Theta(t-s)F_{i,j}(t,s)+\Theta(s-t)F_{j,i}(s,t)
\label{eq:JJ corr formal}
\end{equation}
for arbitrary $t$ and $s$.

\subsection{Fluctuations and the Fick's law}
We now show how to derive \eref{eq:YY} from the correlations of the current.
From \eref{eq:Q evolution} one gets
\begin{equation*}
\llangle Y_i(t)\rrangle=\llangle \mathcal{J}_i(t)\rrangle-\left[ \llangle \tau_i(t)\rrangle-\llangle \tau_{i+1}(t)\rrangle\right]=0
\end{equation*}
and
\begin{eqnarray}
\fl \qquad \llangle Y_i(t)Y_j(s)\rrangle= \llangle Y_i(t)Y_j(s)\rrangle_c=\llangle \mathcal{J}_i(t) \mathcal{J}_j(s) \rrangle_c-\llangle \left[\tau_i(t)-\tau_{i+1}(t)\right]\mathcal{J}_j(s)\rrangle_c\cr
\qquad -\llangle \mathcal{J}_i(t)\left[\tau_j(s)-\tau_{j+1}(s)\right]\rrangle_c +\llangle  \left[\tau_i(t)-\tau_{i+1}(t)\right]\left[\tau_j(s)-\tau_{j+1}(s)\right] \rrangle_c
\label{eq:YiYj step 1}
\end{eqnarray}
For $t\ge s$, the last two terms on the right hand side cancel each other. One writes the second term
\begin{equation*}
\fl \qquad \llangle \left[\tau_i(t)-\tau_{i+1}(t)\right]\mathcal{J}_j(s)\rrangle_c=\frac{\partial}{\partial s}\llangle \left[\tau_i(t)-\tau_{i+1}(t)\right]Q_j(s)\rrangle_c=\frac{d}{ds}K_{i,j}(t,s)=F_{i,j}(t,s)
\end{equation*}
where in the last two steps we used the definition of $K_{i,j}(t,s)$ in \eref{eq:Kij} and the relation \eref{eq:K F rel}.
Substituting in \eref{eq:YiYj step 1} and extending the result for an arbitrary $t$ and $s$ one gets
\begin{equation*}
 \llangle Y_i(t)Y_j(s)\rrangle=\llangle \mathcal{J}_i(t) \mathcal{J}_j(s) \rrangle_c-\Theta(t-s)F_{i,j}(t,s)-\Theta(s-t)F_{j,i}(s,t)
\end{equation*}
At this stage one can use the formula \eref{eq:JJ corr formal} and write
\begin{equation*}
\llangle Y_i(t)Y_j(s)\rrangle=\delta(t-s)\delta_{i,j}\Gamma_{i}(t)
\end{equation*}
where $\Gamma_i(t)$ is given in \eref{eq:D}. This is the correlation for arbitrary system length. In the large $L$ limit, one writes
\begin{equation}
\Gamma_i(t)=\llangle \left[\tau_i(t)-\tau_{i+1}(t)\right]^2\rrangle= 2\llangle \tau_i(t)\rrangle\left[1-\llangle \tau_i(t)\rrangle\right]+\Or\left(\frac{1}{L}\right)
\label{eq:Gamma explicit}
\end{equation}
where we used $\llangle \tau_i(t)\tau_j(t)\rrangle_c\sim \frac{1}{L}$ and $\llangle \tau_{i+1}(t)\rrangle=\llangle \tau_{i}(t)\rrangle+\Or(\frac{
1}{L})$.
This leads to the result \eref{eq:YY}.

\subsection{Steady state}
In the steady state, the equal-time correlation $\llangle \tau_i(t)\tau_j(t)\rrangle_c$ does not change with time. The two-time correlation becomes a function of the time difference and using \eref{eq:tt correlation} one writes
\begin{equation}
\llangle \tau_{i}(t)\tau_{j}(0) \rrangle_c=\sum_{\ell=1}^{L}g_{i,\ell}(t) \llangle \tau_{\ell}\tau_{j} \rrangle_c
\label{eq:tt correlation 2}
\end{equation}

Similarly, in the steady state, $A_{i,j}(t)$ and $\Gamma_i(t)$  are independent of time, leading to 
$F_{i,j}(t,s)\equiv F_{i,j}(t-s)$ in \eref{eq:expression F}.
This makes the integrals in \eref{eq:formal solution} simpler to calculate,
\begin{equation}
\fl  \llangle Q_i(t)Q_j(s)\rrangle_c=s\delta_{i,j}\Gamma_i+\int_{0}^s \!\!\!\! dt^{\prime}(s-t^{\prime})F_{j,i}(t^{\prime})+s\int_{0}^{t-s}\!\!\!\!dt^{\prime}F_{i,j}(t^{\prime})+\int_{t-s}^{t}\!\!\!\!dt^{\prime}(t-t^{\prime})F_{i,j}(t^{\prime})
\label{eq:formal solution two}
\end{equation}
for $t\ge s$.

The correlation $\llangle \mathcal{J}_{i}(t)\mathcal{J}_{j}(s) \rrangle_c$ in the steady state is a function of time difference $t-s$ and from \eref{eq:JJ corr formal}
\begin{equation}
\fl \qquad \llangle \mathcal{J}_{i}(t)\mathcal{J}_{j}(0) \rrangle_c=\delta(t)\delta_{i,j}\Gamma_i+\Theta(t)F_{i,j}(t)+\Theta(-t)F_{j,i}(-t)
\label{eq:JJ corr formal stationary}
\end{equation}

One may compare the results \eref{eq:formal solution two} and \eref{eq:JJ corr formal stationary} with their hydrodynamic formula \eref{eq:qq hydro 0} and \eref{eq:curr corr fhd 2}, respectively. 

\subsubsection*{Explicit formula in the hydrodynamic limit: \label{sec:F large L}}
To write an explicit formula for correlations, one needs to determine the Green's function \eref{eq:Green's function} which is written in terms of eigenfunctions of matrix $\mathbf{M}$ in \eref{eq:matrix M explicit}. 
\begin{equation}
g_{i,j}(t)=\left[e^{-\mathbf{M}t}\right]_{i,j}= \sum_{\lambda}e^{-\lambda t}\psi^{\star}_{i}(\lambda)\psi_{j}(\lambda)
\end{equation}
where $\psi_i(\lambda)$ is the eigenfunction corresponding to eigenvalue $\lambda$, and $\psi^{\star}_i(\lambda)$ is the complex conjugate.
For the real symmetric matrix $\mathbf{M}$ eigenvalues are real; moreover one can verify that eigenvalues are positive.
Then, in the hydrodynamic scale $t=L^2\tau$, for large $L$, the leading contribution comes from the eigenmodes with small $\lambda\; (\sim \frac{1}{L^2})$. For these small eigenvalues, one can verify using $\mathbf{M}$ in \eref{eq:matrix M explicit} that,
\begin{equation*}
\psi_{i}(\lambda)\simeq \sqrt{\frac{2}{L}} \, \sin\left(n\pi\frac{i}{L}\right) \quad \textrm{with}\quad  \lambda\simeq \frac{n^2 \pi^2}{L^2}\qquad \textrm{for large $L$}
\end{equation*}
where $n$ is small positive integer.

The Green's function takes a form
\begin{equation}
g_{i,j}(L^2 \tau)\simeq \frac{2}{L}\sum_{n=1}^{\infty}e^{-n^2\pi^2 \tau}\sin\left(n\pi\frac{i}{L}\right)\sin\left(n\pi\frac{j}{L}\right)
\label{eq:gxy}
\end{equation}
where the upper limit of the summation is set to $\infty$ without altering the leading behavior.

Comparing with \eref{eq:G explicit} one finds that $g_{i,j}(L^2 \tau)\simeq \frac{1}{L}G\left(x,y,\tau\right)$ where the scaled coordinates $(x,y)\equiv\left(\frac{i}{L},\frac{j}{L}\right)$. 

In the large $L$ limit, exact formula of the correlations lead to their hydrodynamic result derived in \sref{sec:fhd}. For the correlation of occupation variables this is easy to see by comparing \eref{eq:c inter} with \eref{eq:tt correlation 2} where one finds
\begin{equation*}
\llangle \tau_i(t)\tau_j(0)\rrangle_c\simeq \llangle \rho(x,\tau)\rho(y,0)\rrangle_c=\frac{1}{L}c(x,y,\tau)
\end{equation*}

Similarly, in the large $L$ limit, $\llangle \mathcal{J}_i(t)\mathcal{J}_j(0)\rrangle_c$ leads to $\llangle J(x,\tau)J(y,0)\rrangle_c$. One easy way to see this is by comparing (\ref{eq:expression F},\ref{eq:Aij}) with their hydrodynamic counterparts (\ref{eq:f inter},\ref{eq:Axy}) for $D(\rho)=1$ and $\sigma(\rho)=2\rho(1-\rho)$. In large $L$ limit, using \eref{eq:Gamma explicit} one gets $\Gamma_j\simeq \sigma(\overline{\rho}(x))$ which then leads to $A_{i,j}\simeq -\frac{1}{L}A(x,y)$ given in \eref{eq:Aij} and \eref{eq:Axy}, respectively. This clearly shows that, $F_{i,j}(t)$ in \eref{eq:expression F} leads to $f(x,y,\tau)$ in \eref{eq:f inter} with a scaling relation 
\begin{equation}
F_{i,j}(t)\simeq \frac{1}{L^3} f\left(x,y,\tau\right) \qquad \textrm{for large $L$}
\label{eq:F scaling}
\end{equation}
As a result, $\llangle \mathcal{J}_i(t)\mathcal{J}_j(0)\rrangle_c$ in \eref{eq:JJ corr formal stationary} and $\llangle J(x,\tau)J(y,0)\rrangle_c$ in \eref{eq:curr corr fhd 2} are related by simple scaling of the current $J(x,\tau)\simeq L\,\mathcal{J}_i(t)$.
One can similarly compare the results for the integrated current $\llangle Q_i(t)Q_j(s)\rrangle_c$ and $\llangle q(x,\tau)q(y,\tau^{\prime})\rrangle_c$ in \eref{eq:formal solution two} and \eref{eq:qq hydro 0}, respectively.

In an alternative approach, one can verify the formula of $f(x,y,\tau)$ in \eref{eq:Fxy} by deriving $F_{i,j}(t)$ for arbitrary system length $L$ and then taking hydrodynamic limit. A derivation is given in \ref{eq:exact derivation}.

\section{Linear response theory \label{sec:linear response}}
In the linear response theory one writes a change in the average value of an observable $B$ due to a small perturbation parameterized by $h(t)$,
\begin{equation}
\llangle \Delta B(t) \rrangle=\int_{-\infty}^{t}ds\, h(s)R(t,s)+\Or(h^2)
\label{eq:Rts}
\end{equation}
where $R(t,s)$ is the response function.

At equilibrium, for a change in the energy of a microscopic configuration $E_c\rightarrow E_c-h(t) A_c$ where $A_c$ is a variable conjugate of the perturbation one finds \cite{Kubo1966,Ruelle2009} a fluctuation dissipation relation
\begin{equation}
R(t,s)= \frac{d \llangle B(t)A(s)\rrangle_c}{ds} \qquad \textrm{for $t\ge s$}
\label{eq:linear response}
\end{equation}
where we have set $k_B T=1$ ($k_B$ is the Boltzmann constant and $T$ is temperature). In \eref{eq:linear response}, $\llangle B(t)A(s)\rrangle_c$ is the unperturbed equilibrium correlation function.

Many recent works  \cite{Seifert2010,Baiesi2009,Baiesi20092,Maes2010,Maes20102,Prost2009,Chetrite2008,Chatelain2003,Diezemann2005,Corberi2007
,Cugliandolo1994,HANGGI1982,Agarwal1972} have given extensions of \eref{eq:linear response} to systems out of equilibrium. In this section, we consider the linear response in the non-equilibrium steady state of the symmetric exclusion process and its generalization to diffusive systems using fluctuating hydrodynamics.

\subsection{Linear response in the symmetric exclusion process}
\begin{figure}
\begin{center}
\includegraphics[width=0.8\textwidth]{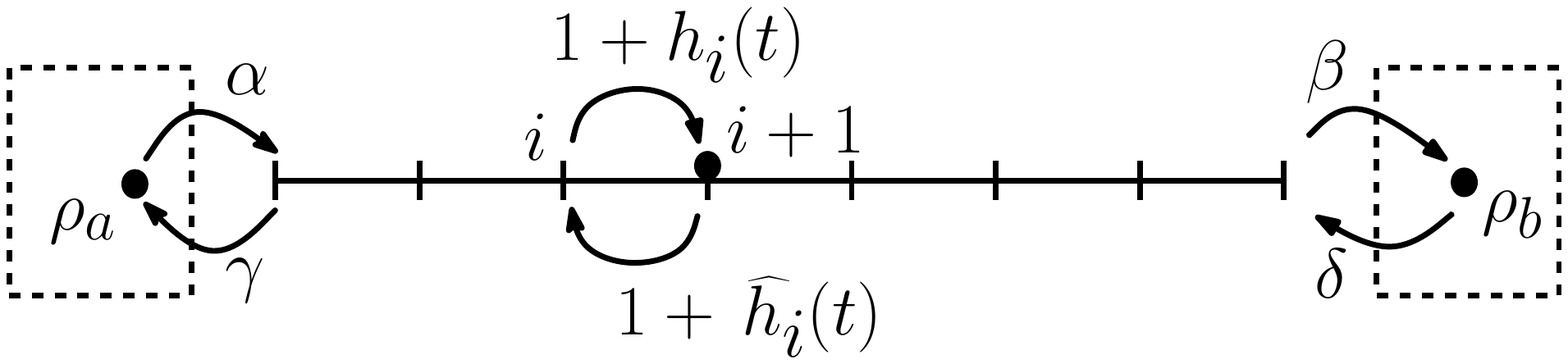}
\end{center}
\caption{Perturbation of jump rates in symmetric exclusion process coupled with reservoirs. \label{fig:perturbation}}
\end{figure}
Let us focus on perturbations of the jump rates at all bonds inside the bulk as shown in \fref{fig:perturbation}.
We consider the effect of this perturbation on the average occupation of a site. The change $\llangle \Delta \tau_i(t)\rrangle$ in the average value of occupation variable $\tau_i(t)$ can be calculated by generalizing \eref{eq:eta evolution} for the perturbed rates and then writing its solution in terms of the Green's function \eref{eq:Green's function}.
Writing $\llangle \Delta \tau_i(t)\rrangle$ in symmetric and antisymmetric parts with respect to the perturbation $h_i(t)$ and $\widehat{h}_i(t)$,
\begin{equation}
\fl \qquad \llangle \Delta \tau_i(t)\rrangle\!\simeq \int_{-\infty}^{t}\!\!\!\!\!\!ds \,\sum_{j=1}^{L-1}\left[\frac{(h_j(s)-\widehat{h}_j(s))}{2}\,R_{1}(i,t;j,s)\!+\!\frac{(h_j(s)+\widehat{h}_j(s))}{2}\,R_{2}(i,t;j,s)\right]
\label{eq:R def sep}
\end{equation}
where we defined two response functions
\numparts
\begin{eqnarray}
 R_{1}(i,t;j,s)= &
\left[g_{i,j+1}(t-s)-g_{i,j}(t-s)\right]\llangle  \left[\tau_j(s)-\tau_{j+1}(s)\right]^2\rrangle
\label{eq:R1} \\
 R_{2}(i,t;j,s)= &
\left[g_{i,j+1}(t-s)-g_{i,j}(t-s)\right]\llangle \tau_{j}(s)-\tau_{j+1}(s)\rrangle. \label{eq:R2}
\end{eqnarray}
\endnumparts

The formulas \eref{eq:R def sep} and (\ref{eq:R1},\,\ref{eq:R2}) give a general result for linear response in the symmetric exclusion process. In spirit of the equilibrium fluctuation dissipation relation one would like to express the response function in terms of two-time correlations. We have achieved this only for the first response function \eref{eq:R1} which can be rewritten as
\begin{equation}
R_1(i,t;j,s)=\frac{d}{ds}\llangle \tau_i(t)Q_j(s)\rrangle_c-\left[\llangle \tau_i(t)\tau_j(s)\rrangle_c-\llangle \tau_i(t)\tau_{j+1}(s)\rrangle_c\right]
\label{eq:R2 final 1}
\end{equation}
This is obtained by using \eref{eq:tQ final} to write
\begin{eqnarray*}
 \frac{d}{ds}\llangle \tau_i(t)Q_j(s)\rrangle_c=\sum_{k}g_{i,k}(t-s)A_{k,j}(s)
\end{eqnarray*}
and then substituting $A_{\ell,j}(s)$ from \eref{eq:Aij} and using \eref{eq:tt correlation} to get
\begin{eqnarray}
\fl  \frac{d}{ds}\llangle \tau_i(t)Q_j(s)\rrangle_c=\llangle \tau_i(t)\tau_j(s)\rrangle_c-\llangle \tau_i(t)\tau_{j+1}(s)\rrangle_c+\left[ g_{i,j+1}(t-s)-g_{i,j}(t-s)\right]\Gamma_j(s)
\label{eq:77}
\end{eqnarray}
Using the expression of $\Gamma_j(s)$ in \eref{eq:D} one can see that the last term in \eref{eq:77} is the $R_1$ in \eref{eq:R1}. This leads to \eref{eq:R2 final 1}.

\subsubsection*{In equilibrium,} the average occupation $\llangle \tau_i \rrangle=\rho$ and substituting in \eref{eq:R2} this leads to $R_2(i,t;j,s)=0$. On the other hand, in the expression \eref{eq:R2 final 1}, using time reversibility one sees that $\llangle \tau_i(t)\tau_j(s)\rrangle_c=\llangle \tau_i(s)\tau_j(t)\rrangle_c$ and  this makes the term inside the square brackets in \eref{eq:R2 final 1} equal to $\frac{d}{dt}\llangle Q_j(t)\tau_i(s)\rrangle_c$. Moreover,
\begin{equation*}
\frac{d}{dt}\llangle Q_j(t)\tau_i(s)\rrangle_c=\llangle \mathcal{J}_j(t)\tau_i(s)\rrangle_c=-\llangle \tau_i(t)\mathcal{J}_j(s)\rrangle_c=-\frac{d}{ds}\llangle \tau_i(t)Q_j(s)\rrangle_c
\end{equation*}
Substituting this in \eref{eq:R2 final 1} one gets 
\begin{equation}
 R_{1}(i,t;j,s)= 2\frac{d}{ds}\llangle \tau_i(t)Q_j(s)\rrangle_c
 \label{eq:fdt Q}
\end{equation}

\subsubsection*{Relation to net escape rate:}
A general formula of the response function in non-equilibrium system has been recently presented in \cite{Baiesi2009,Baiesi20092}. (See \ref{app:linear response} for a derivation in Markov process.) The formula is for a specific type of perturbation which in  \eref{eq:R def sep} corresponds to $\widehat{h}_j(t)=-h_j(t)$; the associated response function is $R_1$.

To write $R_1$ in the form of this general formula one requires an observable: the escape rate from a configuration. We define
\begin{equation*}
\mathcal{E}_j(s,h_j(s))=[1+h_j(s)] \tau_j(s)\left[1-\tau_{j+1}(s)\right]+[1-h_j(s)] \left[1-\tau_j(s)\right]\tau_{j+1}(s)
\end{equation*}
Then $\sum_j\mathcal{E}_j(s,h_j(s))$ is the escape rate from a configuration of occupation variables.
One can verify that \eref{eq:R2 final 1} takes a form 
\begin{equation}
R_1(i,t;j,s)=\frac{d}{ds}\llangle \tau_i(t)Q_j(s)\rrangle_c-\llangle \tau_i(t)\left[\partial_{h} \mathcal{E}_j(s,h)\right]\rrangle_c
\label{eq:R2 final 22}
\end{equation}
which is an explicit example of the general formula obtained in \cite{Baiesi2009,Baiesi20092}.

\subsection{A fluctuating hydrodynamics approach \label{sec:hydro linear}}

We consider a perturbation produced by a small external field $h(x,\tau)$ as in \eref{eq:5}. The perturbation changes average density profile which follows
\begin{equation*}
\partial_{\tau}\overline{\rho}_h(x,\tau)=\partial_x\left[ D(\overline{\rho}_h(x,\tau))\partial_x\overline{\rho}_h(x,\tau)-\sigma(\overline{\rho}_h(x,\tau))h(x,\tau) \right]
\end{equation*}
where the subscript $h$ denotes perturbed state. Writing the difference $u=\overline{\rho}_h-\overline{\rho}$ and considering small $h(x,\tau)$ one gets
\begin{equation*}
\partial_{\tau}u(x,\tau)=\partial^2_x\left[D(\overline{\rho}(x,\tau))u(x,\tau)\right]-\partial_x\left[\sigma(\overline{\rho}(x,\tau))h(x,\tau)\right]
\end{equation*}
Similar to \eref{eq:solution for r} the solution can be written in terms of the Green's function \eref{eq:G},
\begin{equation*}
u(x,\tau)=\int_{-\infty}^{\tau}d\tau^{\prime}\int_0^1 dy\; h(y,\tau^{\prime})\; R_1(x,\tau;y,\tau^{\prime})
\end{equation*}
where
\begin{equation}
R_1(x,\tau;y,\tau^{\prime})=\sigma(\overline{\rho}(y,\tau^{\prime}))\partial_yG(x,y,\tau-\tau^{\prime})
\label{eq:Rq 1}
\end{equation}
This is the hydrodynamic analogue of the microscopic formula \eref{eq:R1} in the symmetric exclusion process.

This response function $R_1(x,\tau;y,\tau^{\prime})$ can be expressed in terms of two-time correlations.
From \eref{eq:solution for r} one can see that
\begin{equation*}
\llangle \rho(x,\tau)\eta(y,\tau^{\prime})\rrangle_c=\llangle r(x,\tau)\eta(y,\tau^{\prime})\rrangle=\frac{1}{L}\sigma(\overline{\rho}(y,\tau^{\prime}))\partial_yG(x,y,\tau-\tau^{\prime}).
\end{equation*}
One can then see from \eref{eq:Rq 1} and \eref{eq:fhd 0} that
\begin{eqnarray*}
\fl \qquad R_1(x,\tau;y, \tau^{\prime}) &=  L \llangle \rho(x,\tau)\eta(y,\tau^{\prime})\rrangle_c \\
&= L \llangle \rho(x,\tau)J(y,\tau^{\prime})\rrangle_c +L \llangle \rho(x,\tau)D(\rho(y,\tau^{\prime}))\partial_y\rho(y,\tau^{\prime})\rrangle_c.
\end{eqnarray*}
To the leading order in large $L$ the formula reduces to
\begin{equation}
\fl \qquad R_1(x,\tau;y, \tau^{\prime}) \simeq L \llangle \rho(x,\tau)J(y,\tau^{\prime})\rrangle_c +L\frac{d }{d y}\left[D(\overline{\rho}(y))\llangle \rho(x,\tau)\rho(y,\tau^{\prime})\rrangle_c\right]
\label{eq:R almost}
\end{equation}
One may compare with the microscopic formula \eref{eq:R2 final 1} in the symmetric exclusion process which corresponds to $D(\rho)=1$.

To obtain \eref{eq:Rq 4 0} one can add
\begin{equation*}
\llangle J(y,\tau)\rho(x,\tau^{\prime})\rrangle_c\simeq -\frac{d}{dy}\left[D(\overline{\rho}(y))\llangle \rho(y,\tau)\rho(x,\tau^{\prime})\rrangle_c\right]
\end{equation*}
to \eref{eq:R almost} and write current $J(y,\tau^{\prime})$ in terms of the integrated current \eref{eq:qJ}.

\subsubsection*{In equilibrium,} formula \eref{eq:Rq 4 0} takes a simple form. Due to the symmetry under time reversal, $\llangle \rho(x,\tau)\rho(y,\tau^{\prime}) \rrangle_c=\llangle \rho(y,\tau)\rho(x,\tau^{\prime}) \rrangle_c$ which makes the last two terms in \eref{eq:Rq 4 0} cancel each other. The first two terms in \eref{eq:Rq 4 0} are equal, $\llangle \rho(x,\tau)J(y,\tau^{\prime})\rrangle_c=-\llangle J(y,\tau)\rho(x,\tau^{\prime})\rrangle_c$ due to symmetry under time reversal. Then the response function
\begin{equation*}
R(x,\tau; y, \tau^{\prime})=2\frac{d}{d\tau^{\prime}}\llangle \rho(x,\tau)q(y,\tau^{\prime})\rrangle_c
\label{eq:Rq eq}
\end{equation*}
The above formula is a generalization of the microscopic result \eref{eq:fdt Q}.

\begin{remark}
The response function $R_2$ in \eref{eq:R2} also has a hydrodynamic analogue. In the fluctuating hydrodynamics this corresponds to a perturbation in the diffusivity, for example, $D_h(\rho(x,\tau))=D(\rho(x,\tau))+h(x,\tau)$. Using an analysis similar for the $R_1(x,\tau;y,\tau^{\prime})$ one gets the corresponding response function
\begin{equation*}
R_2(x,\tau;y,\tau^{\prime})=-\left[\partial_y\overline{\rho}(y,\tau^{\prime})\right]\partial_yG(x,y,\tau-\tau^{\prime})
\end{equation*}
\end{remark}

\section{Summary \label{sec:summary}}
In this work, we have used fluctuating hydrodynamics to calculate the two-time correlations (\ref{eq:rhorho hydro},\,\ref{eq:c}) of the density and (\ref{eq:curr corr fhd 2},\,\ref{eq:f macro}) of the current in the out-of-equilibrium steady state of diffusive system. This allowed us to obtain the spectral distribution of the current (\ref{eq:Sxn},\,\ref{eq:S0}) and to obtain correlations of the integrated current \eref{eq:qq hydro 0}. We have also discussed an extension of the linear response theory in non-equilibrium steady state where the response function is expressed in terms of two-time correlations. Our approach applies to arbitrary diffusive systems of which the symmetric exclusion process is a special case. The hydrodynamics expressions are consistent with results obtained from an explicit microscopic calculation in the symmetric simple exclusion process.

It would be interesting to compare our result of the spectral distribution of the current with experiments in quasi-one dimensional diffusive systems, like, transport inside channels in porous medium \cite{karger1992}.

The two-time correlations are expressed in terms of Green's function. A straightforward generalization of our method shows that higher order correlations can also be expressed in terms of the same Green's function; it would be intriguing to find an explicit formula of all higher order correlations, in particular, their cumulant generating function in terms of the Green's function.

The fluctuating hydrodynamics approach can be generalized to more complicated systems (higher dimensions, several species of particles, multiple-conserved quantities, presence of external field). More challenging would be to extend it to \textit{non-linear} fluctuating hydrodynamics which have been recently developed \cite{Spohn2016} to explain anomalous transport in one-dimensional Hamiltonian system.

\ack
The work of TS is supported by a junior research chair of the Philippe Meyer Institute for Theoretical Physics at Ecole Normale Sup\'{e}rieure, Paris. The authors would like to acknowledge fruitful discussions with Lyd\'{e}ric Bocquet, David Mukamel, Fr\'{e}d\'{e}ric van Wijland, and Francesco Zamponi.

\appendix

\section{Derivation of \eref{eq:f+f id} \label{sec:f+f id}} From \eref{eq:c} and \eref{eq:f macro} one can see that
\begin{eqnarray*}
\fl \int_0^{\infty}ds\left[f(y,x,s)+f(x,y,s)\right]=-\sigma(\overline{\rho}(y))\frac{d^2U(x,y)}{dxdy}-\sigma(\overline{\rho}(x))\frac{d^2U(y,x)}{dydx}\cr
\qquad \qquad \qquad \qquad \qquad \qquad +\int_{0}^{1}dz\sigma(\overline{\rho}(z))\frac{d^2U(x,z)}{dxdz}\frac{d^2U(y,z)}{dydz}
\end{eqnarray*}
where we defined
\begin{equation}
U(x,y)=D(\overline{\rho}(x))\int_{0}^{\infty}ds G(x,y,s)
\label{eq:U}
\end{equation}
Using \eref{eq:G} one finds $\partial^2_{x}U(x,y)=-\delta(x-y)$ along with the boundary condition $U(0,y)=U(1,y)=0$. The solution is
\begin{equation}
U(x,y)=x(1-y)\Theta(y-x)+y(1-x)\Theta(x-y)
\label{eq:U explicit}
\end{equation}
Substituting the explicit formula of $U(x,y)$ one gets \eref{eq:f+f id}.

\section{Derivation of \eref{eq:QiQjss} \label{app:derivation of A}}
The integrated current $Q_i(t)$ is the net number of particles transfered from site $i$ to $i+1$ in a time window $0$ to $t$. In the bulk, a change in $Q_i(t)$ in an infinitesimal time interval $dt$ is given by
\begin{equation*}
\fl  Q_i(t+dt)=\cases{Q_i(t)+1 & with probability $\tau_i(t)\left[1-\tau_{i+1}(t)\right]dt$ \cr
Q_i(t)-1 & with probability $\left[1-\tau_{i}(t)\right]\tau_{i+1}(t)dt$ \cr
Q_i(t) & with probability $1\!-\!\left[\tau_i(t)\!+\!\tau_{i+1}(t)\!-\!2\tau_i(t)\tau_{i+1}(t)\right]dt$
}
\label{eq:rates Q}
\end{equation*}
Using this one gets
\begin{equation}
 \frac{d}{dt}\!\! \llangle Q_{i}(t)Q_{j}(t)\rrangle_c\!=K_{i,j}(t,t)+K_{j,i}(t,t)+\delta_{i,j}\Gamma_j(t)
 \label{eq:QQtt solution}
\end{equation}
where the quantities $K_{i,j}(t,s)$ and $\Gamma_i(t)$ are given in \eref{eq:Kij} and \eref{eq:D}, respectively.
Integrating \eref{eq:QQtt solution} over time one gets \eref{eq:QiQjss}.

\section{An exact result for the symmetric exclusion process of arbitrary length \label{eq:exact derivation}}
The analysis is simple to present for a choice of rates at the boundary such that $a$ and $b$ in \eref{eq:ab} are both equal to $1$. In the large $L$ limit, correlations do not explicitly depend on $(a,b)$ and one gets the same hydrodynamic result as derived in \sref{sec:sep explicit}.

For $(a,b)\equiv (1,1)$ the microscopic Green's function \eref{eq:Green's function} takes a form
\begin{equation}
g_{i,j}(t)=\left[e^{-\mathbf{M}t}\right]_{i,j}=\frac{2}{L+1}\sum_{n=1}^{L}e^{-\lambda_n t}\sin(\theta_n i)\sin(\theta_n j)
\label{eq:gij exact}
\end{equation}
where we defined
\begin{equation*}
\lambda_n=2\left(1-\cos\theta_n\right)\qquad \textrm{and}\qquad \theta_n=\frac{n \pi}{L+1}
\end{equation*}
One can verify this using matrix $\mathbf{M}$ in \eref{eq:matrix M explicit}.

To calculate $\llangle \tau_i(t)\tau_j(0)\rrangle_c$ in \eref{eq:tt correlation 2} one needs to use the result \eref{eq:stationary tautau} for the steady state correlation at equal-time.
This leads to an explicit formula
\begin{eqnarray}
\fl \llangle \tau_i(t)\tau_j(0)\rrangle_c\!=\!\!\sum_{n=1}^{L}\frac{2e^{-\lambda_n t}}{L+1}\!\sin\!\left(\theta_n i\right)\!\sin\!\left(\theta_n j\right)\!\left\{\!\llangle\tau_j\rrangle\!(1\!-\!\llangle\tau_j\rrangle)\!\!+\!\!\frac{(\rho_a\!\!-\!\!\rho_b)^2}{L(L+1)}\!\!\left[\frac{j(L\!\!+\!\!1\!\!-\!\!j)}{L+1}\!-\!\frac{1}{\lambda_n}\right]\!\right\}
\label{eq:tautau micro original}
\end{eqnarray}
In the large $L$ limit this leads to the asymptotic result \eref{eq:tautau macro 0}.

To do a similar calculation for the correlation of current \eref{eq:JJ corr formal stationary} one uses \eref{eq:gij exact} into \eref{eq:expression F} and gets
\begin{equation}
\fl \qquad F_{i,j}(t)=-\frac{4}{L+1}\sum_{n=1}^{L}e^{-\lambda_n t}\sin\left(\frac{\theta_n}{2}\right)\cos\left(\frac{\theta_n}{2}(2i+1)\right)\sum_{\ell}\sin\left(\theta_n \ell \right)A_{\ell,j}
\label{eq:F intermediate}
\end{equation}

The quantity $A_{i,j}$ is written in \eref{eq:Aij} in terms of the average occupation $\llangle \tau_i \rrangle$ and the correlation $\llangle \tau_i\tau_j \rrangle_c$ which are given in (\ref{eq:average results},\ref{eq:stationary tautau}), respectively. Substituting in \eref{eq:F intermediate} one gets an explicit expression of $F_{i,j}(t)$, and the correlation of current is obtained by substituting this formula in \eref{eq:JJ corr formal stationary}.

The spectral power density of current across bond $(i,i+1)$ is given by
\begin{eqnarray*}
\fl \qquad S_i(\omega)&=\int_{-\infty}^{\infty}dt\,e^{-\hat{i}\,2\pi\,\omega \,t}\llangle \mathcal{J}_i(t)\mathcal{J}_i(0) \rrangle_c=\Gamma_i+2\int_0^{\infty}dt\cos(2\pi \omega t)F_{i,i}(t)
\label{eq:Si}
\end{eqnarray*}
where we used \eref{eq:JJ corr formal stationary}. One can derive explicit formula using \eref{eq:F intermediate}. For example, at the right boundary, one gets
\begin{eqnarray*}
\fl S_L(\omega)=\frac{1}{L+1}\left\{\rho_a+\rho_b-2\rho_a\rho_b+\rho_b(1-\rho_b)\sum_{n=1}^{L}\frac{4\pi^2\omega^2\left(\cos\frac{\theta_n}{2}\right)^2}{4\left(\sin\frac{\theta_n}{2}\right)^4+\pi^2\omega^2}\right. \cr \qquad \qquad \qquad \qquad \qquad \qquad \qquad \left.-\frac{(\rho_a-\rho_b)^2}{L(L+1)}\sum_{n=1}^{L}\frac{\left(\sin\theta_n\right)^2}{4\left(\sin\frac{\theta_n}{2}\right)^4+\pi^2\omega^2}\right\}
\label{eq:SL final}
\end{eqnarray*}
The expression \eref{eq:S0 exact} at large $L$ is obtained by using
\begin{equation*}
S(1,\nu)\simeq S_L\left(\frac{\nu}{L^2}\right)
\end{equation*}

\section{Linear response theory for a general Markov process \label{app:linear response}}
There are several ways of writing the response function in out of equilibrium systems \cite{Seifert2010,Baiesi2009,Baiesi20092,Maes2010,Maes20102,Prost2009,Chetrite2008,Chatelain2003,Diezemann2005,Corberi2007
,Cugliandolo1994,HANGGI1982,Agarwal1972}. In a recent work \cite{Baiesi2009} a connection has been made between the response function and the net escape rate from a configuration. We present here an elementary derivation for a discrete Markov process.

Consider a Markov process with transition rates $\omega(c^{\prime},c)$ from configuration $c$ to configuration $c^{\prime}$. The probability $P(c,t)$ of a configuration $c$ at time $t$ follows
\begin{equation}
\frac{dP(c,t)}{dt}=\sum_{c^{\prime}}\left[\omega(c,c^{\prime})P(c^{\prime},t)-\omega(c^{\prime},c)P(c,t)\right]
\label{eq:rate eqn}
\end{equation}
One can define the Green's function associated to the process as the solution of
\begin{equation}
\fl \qquad\qquad  \frac{d}{dt}\mathcal{G}_{c,c^{\prime}}(t)=\sum_{c^{\prime\prime}}\left[\omega(c,c^{\prime\prime})\mathcal{G}_{c^{\prime\prime},c^{\prime}}(t)-\omega(c^{\prime\prime},c)\mathcal{G}_{c,c^{\prime}}(t)\right]\quad \textrm{and}~  \mathcal{G}_{c,c^{\prime}}(0)=\delta_{c,c^{\prime}}
\label{eq:G cal}
\end{equation}
One can verify that this Green's function satisfies also
\begin{equation}
\frac{d}{dt}\mathcal{G}_{c,c^{\prime}}(t)=\sum_{c^{\prime\prime}}\left[\mathcal{G}_{c,c^{\prime\prime}}(t)-\mathcal{G}_{c,c^{\prime}}(t)\right]\omega(c^{\prime\prime},c^{\prime})
\label{app:g reversed}
\end{equation}

A general perturbation of the transition rates can be written as
\begin{equation}
\omega_{h}(c,c^{\prime})=\omega(c,c^{\prime})e^{[h(t)H(c,c^{\prime})+\Or(h^2)]} \qquad \textrm{for $c\ne c^{\prime}$}
\label{app:H}
\end{equation}
where $h(t)$ is a small time dependent parameter, and $H(c,c^{\prime})$ is arbitrary.
A change in the average of an observable $B_c$ is given by
\begin{equation}
\fl \qquad \qquad  \llangle \Delta B(t)\rrangle=\sum_cB_c\left[P_h(c,t)-P(c,t)\right]=\int_{-\infty}^{t}ds h(s)R(t,s)+\Or(h^2)
\label{eq:R def Markov}
\end{equation}
where $R(t,s)$ is the response function, and $P_h(c,t)$ is the probability distribution of the Markov process with the perturbed rate $\omega_h(c,c^{\prime})$.

Using \eref{eq:rate eqn} and a similar rate equation for $P_h(c,t)$ one can verify that
\begin{equation*}
\fl P_h(c,t)-P(c,t)\simeq \int_{-\infty}^{t}ds\,h(s)\sum_{c^{\prime},c^{\prime\prime}}\left[\mathcal{G}_{c,c^{\prime}}(t-s)-\mathcal{G}_{c,c^{\prime\prime}}(t-s)\right]H(c^{\prime},c^{\prime\prime})\omega(c^{\prime},c^{\prime\prime})P(c^{\prime\prime},s)
\end{equation*}
up to linear order in $h(s)$.
Comparing with \eref{eq:R def Markov} one can see that
\begin{equation}
\fl \qquad R(t,s)=\sum_{c\,, c^{\prime}, c^{\prime\prime}}B_c\left[\mathcal{G}_{c,c^{\prime}}(t-s)-\mathcal{G}_{c,c^{\prime\prime}}(t-s)\right]H(c^{\prime},c^{\prime\prime})\omega(c^{\prime},c^{\prime\prime})P(c^{\prime\prime},s)
\label{app:R general}
\end{equation}
This is the response function for a general Markov process.

\subsection*{Relation to the escape rate}
The formula \eref{app:R general} can be rewritten as $R=R_A+R_S$ where
\begin{eqnarray*}
R_A(t,s)&=&\sum_{c,c^{\prime}}\sum_{c^{\prime\prime}\ne c^{\prime}}B_c\, \mathcal{G}_{c,c^{\prime}}(t-s)H(c^{\prime},c^{\prime\prime})\omega(c^{\prime},c^{\prime\prime})P(c^{\prime\prime},s) \\
R_S(t,s)&=&-\sum_{c,c^{\prime}}B_c\, \mathcal{G}_{c,c^{\prime}}(t-s)\left[\sum_{c^{\prime\prime}\ne c^{\prime}}H(c^{\prime\prime},c^{\prime})\omega(c^{\prime\prime},c^{\prime})\right]P(c^{\prime},s)
\end{eqnarray*}

The second part is a two-time correlation 
\begin{equation}
R_S(t,s)=-\llangle B(t)a(s)\rrangle
\label{eq:RS}
\end{equation}
where
\begin{equation}
a_{c^{\prime}}=\sum_{c^{\prime\prime}\ne c^{\prime}}H(c^{\prime\prime},c^{\prime})\omega(c^{\prime\prime},c^{\prime})
\label{eq:a def}
\end{equation}
The observable $a_c$ has an interpretation in terms of the escape rate
$\mathcal{E}_{h}(c)=\sum_{c^{\prime}\ne c}\omega_{h}(c^{\prime},c)$.
Using \eref{app:H} and \eref{eq:a def} one can see that $a_c=\partial_{h}\mathcal{E}_{h}(c)\big\vert_{h=0}$.

\subsection*{A special perturbation}
The term $R_A$ can however be expressed in terms of correlation function, for a perturbation \eref{app:H} with $H(c^{\prime},c)=V_{c^{\prime}}-V_c$ where $V_c$ is an arbitrary function. This can be done using
\begin{equation*}
\fl \frac{d}{ds}\llangle B(t)V(s)\rrangle=\sum_{c, c^{\prime\prime}}\left[B_c\frac{d}{ds}\mathcal{G}_{c,c^{\prime\prime}}(t-s)V_{c^{\prime\prime}}P(c^{\prime\prime},s)+B_c\,\mathcal{G}_{c,c^{\prime\prime}}(t-s)V_{c^{\prime\prime}}\,\frac{d}{ds}P(c^{\prime\prime},s)\right]
\end{equation*}
and subsequently simplifying the expression using \eref{app:g reversed} one gets
\begin{equation*}
R_A(t,s)=\frac{d}{ds}\llangle B(t)V(s)\rrangle
\end{equation*}

Using together with \eref{eq:RS} one gets the response function
\begin{equation}
R(t,s)=R_A(t,s)+R_S(t,s)=\frac{d}{ds}\llangle B(t)V(s)\rrangle-\llangle B(t)a(s)\rrangle
\label{eq:Rts int}
\end{equation}
One can write the response function in an alternative form:
\begin{equation}
\fl \qquad R(t,s)=\left[\frac{d}{ds}\llangle B(t)V(s)\rrangle-\frac{d}{dt}\llangle V(t)B(s)\rrangle\right]+\left[\llangle a(t)B(s)\rrangle-\llangle B(t)a(s)\rrangle\right]
\label{app:R final}
\end{equation}
which is obtained by adding
\begin{equation*}
\llangle a(t)B(s)\rrangle-\frac{d}{dt}\llangle V(t)B(s)\rrangle=0
\end{equation*}
to \eref{eq:Rts int} which is easy to derive using \eref{eq:a def} with $H(c^{\prime},c)=V_{c^{\prime}}-V_c$ and \eref{eq:G cal}.

Under this form \eref{app:R final} it is easy to see that at equilibrium 
\begin{equation*}
R(t,s)=2\frac{d}{ds}\llangle B(t)V(s)\rrangle
\end{equation*}
Due to time reversal symmetry, the two terms with $a_c$ in \eref{app:R final} cancel each other, whereas the first two terms with $V_c$ are equal, leading to equilibrium fluctuation dissipation relation

Formulas  \eref{eq:Rts int}-\eref{app:R final} were originally derived in \cite{Baiesi2009,Baiesi20092}. Result for more general perturbation can be found in \cite{Maes2010,Maes20102,Diezemann2005}. 

\begin{remark}
The microscopic formula \eref{eq:R2 final 1} can be related to \eref{eq:Rts int} by considering the symmetric exclusion process as a Markov process on a joint configuration of occupation variables and integrated current $c\equiv \{\tau_i, Q_i\}$. The formula \eref{eq:R2 final 1} is for a perturbation of jump rates (see \fref{fig:perturbation}) with $\widehat{h}_i(t)=-h_i(t)$. In the Markov process this translates into a change in the transition rates
\begin{equation}
\omega_h(c^{\prime}, c)=\omega(c^{\prime},c)e^{\sum_ih_i(t)\left[Q_i(c^{\prime})-Q_i(c)\right]}
\label{eq:tr}
\end{equation}
equivalently $V(c)\equiv Q_i(c)$. Substituting this in \eref{eq:Rts int} for observable $B(c)\equiv \tau_i$ and calculating $a_c$ using \eref{eq:tr} one gets \eref{eq:R2 final 1}.
\end{remark}

\section*{References}
\bibliographystyle{iopart-num}
\bibliography{reference}

\end{document}